\documentclass[11pt]{article}
\usepackage{amsfonts}
\usepackage{amsbsy}
\usepackage{amscd}

\def\be{\begin{equation}}
\def\ee{\end{equation}}
\def\bea{\begin{eqnarray}}
\def\eea{\end{eqnarray}}

\def\R{{\mathbb R}}
\def\1{\'{\i}}

\def\qq{\hat p}

\def\dd{{\rm d}}

\def\k{\omega}
\def\>#1{{\bf #1}}                 

\def\ro{\rho}
\def\s{s}
\def\yy{y}
\def\et{\eta}

\begin{document}

\hfill\ 
\bigskip

\begin{center}
{\Large{\bf{Non-commutative relativistic  spacetimes and  
worldlines }}}\\[6pt]
 
{\Large{\bf{from  2+1 quantum (anti-)de Sitter  groups}}}
 
\bigskip

{\Large{\bf{}}}

\end{center}

\bigskip

\begin{center}   {\sc \'Angel Ballesteros, N.~Rossano Bruno  and
Francisco~J.~Herranz }
\end{center}

\begin{center} {\it { 
 Departamento de F\1sica,  Universidad de Burgos, E- 09001 Burgos, Spain}}\\[4pt] 
 e-mail:
angelb@ubu.es,  r.bruno@bnrenergia.it, fjherranz@ubu.es
\end{center}

\bigskip\bigskip

\begin{abstract} 
\noindent
The $\kappa$-deformation of the (2+1)D  anti-de Sitter,
Poincar\'e and de Sitter  groups  is presented through a unified approach in which the curvature
of the spacetime (or the cosmological constant) is considered as an explicit parameter.
 The  Drinfel'd--double and the Poisson--Lie structure underlying
the $\kappa$-deformation are explicitly given, and the three quantum kinematical groups are
obtained as quantizations of such Poisson--Lie algebras. As a consequence, the 
non-commutative  (2+1)D  spacetimes that  generalize   the $\kappa$-Minkowski space to the (anti-)de
Sitter ones are obtained. Moreover, noncommutative 4D spaces of  (time-like) geodesics  can be defined,  and they can be interpreted as a novel possibility to introduce 
non-commutative worldlines. Furthermore, quantum (anti-)de Sitter algebras are presented both 
in the known basis related with 2+1  quantum gravity   and    in a new    one which
generalizes the bicrossproduct  one. In this framework, the quantum deformation parameter is
related with the Planck length, and the existence of a
kind of  ``duality" between the cosmological constant and the Planck scale is also
envisaged.
\end{abstract}

\bigskip\medskip 
\bigskip\medskip



\newpage 


\section{Introduction}

The  connection between quantum groups and Planck scale physics was early suggested in~\cite{Majida}.  Quantum deformations of Lie algebras and Lie groups~\cite{KR,drinfeld87, Jimbo, Tak, FRT, CP, Majid} have
been broadly applied in the construction of deformed  symmetries of
spacetimes~\cite{LukierskiRuegg1992,CGh1,Ita,BCGpho,Giller,Lukierskib,Maslanka,Majid:1994cy,Zak,LukR,CK3,CK4,Null, LukNR, Tmatrix}, especially for the Poincar\'e and Galilei  cases, for which the deformation parameter
is known to play the role of a fundamental scale. Amongst all these quantum kinematical
algebras the well known $\kappa$-Poincar\'e algebra~\cite{LukierskiRuegg1992,Giller,Lukierskib,Majid:1994cy,LukR} has been frequently considered.

These deformed Poincar\'e  symmetries were later applied in the context of the so called doubly special relativity (DSR)
theories~\cite{Amelino-Camelia:2000mrr,Amelino-Camelia:2000mn, 
Amelino-Camelia:2002vy,MagueijoSmolin,Kowalski-Glikman:2002we,Kowalskia,Kowalskib,Lukierski:2002df,DSRjpa} 
which introduced
 two fundamental scales:  the usual observer-independent velocity scale $c$ as well as an
observer-independent length scale $l_p$, which was related to the deformation parameter in the algebra. Since
from all approaches to  quantum gravity~\cite{Polchinski:1996na,Carlip:2001wq,Smolin:2002sz,
Forste:2001ah,Freidel:2002hx} the Planck scale is thought to play a
fundamental role, DSR theories seem to establish a promising link between some Planck scale effects and quantum groups~\cite{amel,KowalskiFS}.    

From a more general viewpoint,   we recall that non-commutative spaces have been proposed as a suitable algebraic framework in order  to  describe the ``quantum" structure of the geometry of spacetime at the Planck scale through a non-commutative algebra
of  quantum spacetime coordinates~\cite{Snyder, Woronowicz, FredenCMP, Szabo, Verlinde}. In this way the deformation parameter characterizes  the non-commutativity of the spacetime algebra, thus  generating uncertainty relations between non-commuting coordinates that can   be  thought to model a  ``fuzzy" or ``discrete" nature of the spacetime at very small distances (or high energies)~\cite{Maggiore, Garay}. In particular, the non-commutative spacetime deduced from the $\kappa$-Poincar\'e algebra is the so-called $\kappa$-Minkowski spacetime~\cite{
Maslanka,LukNR},
which is the algebra defined by the spacetime quantum group coordinates dual to the translation (momenta)
generators.

In this framework, spacetime curvature (or non-zero cosmological constant) should play a relevant role concerning the possible cosmological consequences of a quantum spacetime (see, e.g.,~\cite{amel,Brunoc,  Marciano, iceCUBE, Giulia1,Giulia2} and references therein).   
Therefore, it seems natural to consider the construction of the
$\kappa$-deformation for the (anti-)de Sitter (hereafter (A)dS)  groups, and  to analyse their possible connections with quantum gravity theories with a non-zero cosmological constant. In this
respect, we recall that the Hopf structure for the $\kappa$-deformation of the
(2+1)D  (A)dS  and Poincar\'e ($\cal P$) algebras were collectively obtained
in~\cite{CK3}, and their connection between their deformed commutation rules and 2+1
quantum gravity has been   explored in~\cite{amel}. The results obtained
in~\cite{CK3} correspond to the l.h.s.\ of the commutative diagram:
\be
\CD 
AdS @>{z} >>  U_z(AdS)  \\
@A{\k\,=+\frac 1 {R^2}  }AA  
@A{\k\ }  AA\\ 
{\cal P} @>{z} >>  U_z({\cal P}) \\ 
@V{\k\,=-\frac 1 {R^2}  }VV  
@V{\k\ }VV\\ 
dS @>{z} >>  U_z(dS) 
\endCD
\CD 
\quad @>{\rm duality }>>\quad   
\endCD
\CD 
{\rm Fun}(AdS) @>{z} >>  {\rm Fun}_z(AdS)  \supset {\bf AdS}_z^{2+1} \\
@A{\k\ }AA  
@A{\k\ }  AA\\ 
{\rm Fun}({\cal P}) @>{z} >>  {\rm Fun}_z({\cal P}) \supset 
{\bf M}_z^{2+1}\\  @V{\k\ }VV  
@V{\k\ }VV \\ 
 {\rm Fun}(dS) @>{z} >>  {\rm Fun}_z(dS) \supset   {\bf dS}_z^{2+1}
\endCD
\label{aa}
\ee
where vertical arrows indicate a classical deformation~\cite{Ober,Ober2} that introduces the spacetime curvature
$\k$ (or cosmological constant $\Lambda= -\k$) related to the (A)dS radius $R$  
by $\k=\pm 1/R^2$, and the horizontal ones show the quantum deformation with parameter
$z=1/\kappa$  (related to the Planck length $l_p$);  reversed arrows correspond 
to the spacetime contraction
$\k\to 0$ and (classical) non-deformed limit $z\to 0$.
As a consequence, the construction of non-commutative  (A)dS  spacetimes  in
terms of intrinsic and ambient spacetime quantum group coordinates seems worth to be explored in detail and, moreover, 
the same framework could account for new proposals
of non-commutative spaces of time-like geodesics (worldlines), which --to the best of our knowledge-- have not been considered in the literature yet, even for the Poincar\'e case. 

Here we present an enlarged and updated review version of our unpublished manuscript arXiv:hep-th/0401244,     
in which the above mentioned problems are faced for the three relativistic
cases simultaneously, that is, by dealing explicitly with the spacetime curvature $\k$ as a contraction parameter. 
Hence, we propose to explore the r.h.s.\ of the diagram
(\ref{aa}) (dual to the l.h.s.)  by   computing the quantum deformation of the (2+1)D  (A)dS
groups (that is,~${\rm{Fun}}_z((A)dS)$) which are  obtained by quantizing the   Poisson--Lie
algebra of smooth functions on these groups (namely,~Fun($(A)dS)$) coming from a suitable classical
$r$-matrix.  
In this way, the (2+1)D non-commutative spaces (e.g.~${\bf AdS}_z^{2+1}$) can then be
identified as certain subalgebras of the corresponding quantum groups.  Moreover, we also construct and study in detail the corresponding 4D   non-commutative spaces of worldlines.   We   stress that in our
approach  all the $\kappa$-Poincar\'e relations (including   its non-commutative spaces)
can be directly recovered from the general (A)dS expressions through the 
limit $\k\to 0$. Moreover, all of the resulting non-commutative spaces are covariant under quantum group (co)actions (for the construction of Poisson and quantum homogeneous spaces we refer to~\cite{DrHS, Koor, Zakrzewski, Reyman, Ciccoliqplanes} and references therein).

The structure of the paper is the following. 
In the next section we recall the basics on the (A)dS  groups in (2+1) dimensions
and their associated homogeneous (2+1)D spacetimes and 4D spaces of worldlines
(time-like lines).   Both types of spaces are described in terms of intrinsic
quantities (related to group parameters) as well as   in ambient coordinates
 with one and two extra dimensions, respectively, which will be further
used in their non-commutative versions.
 By starting from the classical $r$-matrix that generates the $\kappa$-deformation, we
construct in section 3  the corresponding Drinfel'd-double and  obtain  some preliminary information on
the first-order quantum deformation, from which first-order  non-commutative spaces arise. On one hand, we find that at first-order in the deformation parameter the three 
non-commutative relativistic spacetimes are given by the same $\kappa$-Minkowski algebra. Moreover, we show that the deformation
parameter can be interpreted as a curvature on a classical dS spacetime for the three
cases, thus generalizing the results obtained in~\cite{Kowalskia,Kowalskib} for 
$\kappa$-Poincar\'e. On the other hand, we obtain that the first-order  non-commutative
spaces of worldlines are in fact  non-deformed ones,  and a relationship with the
non-relativistic (Newtonian) kinematical groups is thus established.

As an intermediate stage in the search of  the  quantum (A)dS  groups, we
compute in section 4 the invariant (A)dS vector fields and next the Poisson--Lie
structures coming from the classical $r$-matrix generating the $\kappa$-deformation. These results enable us to  propose in section 5
the non-commutative (A)dS spaces, which  are written in both intrinsic and
ambient coordinates.   The resulting non-commutative spacetimes    show how the
curvature modifies the underlying first-order $\kappa$-Minkowski space, while for the
non-commutative spaces of worldlines we find that 2D velocity/rapidity space (spanned by the dual coordinates to the boost generators) remains
non-deformed for $\kappa$-Poincar\'e but becomes deformed for the (A)dS cases.
Hence   Lorentz invariance seems to be lost (or somewhat ``deformed") when a non-zero
curvature/cosmological constant  is considered.

  Section 6 is devoted to study the (dual)
quantum (A)dS algebras and their deformed Casimirs in two different basis. In
particular, starting from the expressions given in~\cite{CK3},
a non-linear transformation involving the generators of the stabilizer subgroup of a
time-like line   allows us to obtain these quantum algebras in a new
basis that generalizes for any $\k$  the bicrossproduct basis of
$\kappa$-Poincar\'e~\cite{Majid:1994cy}. These results are analysed in connection
with     2+1 quantum gravity~\cite{amel} and a
``duality" between curvature/cosmological constant and deformation parameter/Planck
length is suggested along the same lines of the so-called ``semidualization" approach for Hopf algebras in 2+1 quantum gravity~\cite{MSsemid} associated to the exchange of the cosmological length scale and the Planck mass (see also~\cite{OSsemid, Oseir}).   Finally, some remarks and comments  concerning recent findings in this framework  close the paper.


\section{(Anti-)de Sitter Lie groups and their homogeneous spaces}

The Lie algebras of the three (2+1)D relativistic spacetimes of constant
curvature 
  can    collectively be described by means of a real (graded) contraction
parameter $\k$ \cite{CK3},   and we denote them by $so_\k(2,2)$.
If $\{J,P_0, \>P=(P_1,P_2), \>K=(K_1,K_2)\}$   are, in this order,
the generators of rotations, time translations, space translations and
boosts,   the commutation relations of  $so_\k(2,2)$ read
 \be  
\begin{array}{lll} 
[J,P_i]=   \epsilon_{ij}P_j  ,&\qquad
[J,K_i]=   \epsilon_{ij}K_j  ,&\qquad  [J,P_0]= 0 ,  \\[2pt]
[P_i,K_j]=-\delta_{ij}P_0 ,&\qquad [P_0,K_i]=-P_i ,&\qquad
[K_1,K_2]= -J   , \\[2pt]
[P_0,P_i]=\k K_i ,&\qquad [P_1,P_2]= -\k J  ,
\end{array}
\label{ba}
\ee 
where  from now on we assume that Latin indices $i,j=1,2$, Greek
ones $\mu,\nu=0,1,2$,  $\hbar = c =1$ and
$\epsilon_{ij}$ is a skewsymmetric tensor such that $\epsilon_{12}=1$. 
For a positive, zero and negative value of $\k$,   $so_\k(2,2)$  provides a Lie algebra isomorphic to
$so(2,2)$, $iso(2,1)$ and $so(3,1)$, respectively. 
 The case  $\k= 0$ can also be  understood as an In\"on\"u--Wigner
contraction~\cite{IWa}:  $so(2,2)\to iso(2,1)\leftarrow so(3,1)$.

Parity $\Pi$ and time-reversal $\Theta$ are involutive automorphisms of 
$so_\k(2,2)$  defined by \cite{BLL}:
\be 
\begin{array}{ll} 
\Pi  :& (P_0, \>P , \>K ,J)\to  (P_0,- \>P ,- \>K ,J) , \\[2pt]
\Theta  :& (P_0, \>P , \>K ,J)\to  (-P_0, \>P ,- \>K ,J)  ,
\end{array}
\label{bb}
\ee 
which together with the  composition 
\be
\Pi\Theta  :\ (P_0, \>P , \>K ,J)\to  (-P_0,- \>P , \>K ,J) ,
\label{comp}
\ee
 and the identity determine
a  $\mathbb Z_2\otimes \mathbb Z_2\otimes  \mathbb Z_2$ Abelian group of
involutions~\cite{graded}. 
The  automorphisms $\Pi\Theta$ and $\Pi$
 give rise, in this order, to the  Cartan
decompositions:
\bea
&&\!\!\!\!\!\!\!\!\!\!\!\!  {so}_{\k}(2,2)={ {h}_{(1)}}\oplus   {p}_{(1)},\qquad  
{ {h}_{(1)}}=\langle  \>K ,J \rangle\simeq  {so} (2,1),\qquad 
{ {p}_{(1)}}=\langle  P_0, \>P \rangle ,  \label{spacea}\\[2pt]
&&\!\!\!\!\!\!\!\!\!\!\!\!  {so}_{\k}(2,2)={ {h}_{(2)}}\oplus   {p}_{(2)},\qquad  
{ {h}_{(2)}}=\langle J,P_0  \rangle\simeq  {so} (2) \oplus  {so}_\k (2),\qquad 
{ {p}_{(2)}}=\langle    \>P, \>K \rangle ,\label{spaceb}
\eea
where ${ {h}_{(1)}}\simeq {so} (2,1)$ is the Lorentz subalgebra and $ {so}_\k (2)= \langle P_0  \rangle $ covers the Lie subalgebras $ {so} (2)$, $i  {so}  (1)\simeq \mathbb R$  and  $ {so} (1,1)$ for $\k>,=,<0$, correspondingly.  In fact, the
contraction parameter $\k$ is related to   the $\mathbb Z_2$-grading associated to
$\Pi\Theta$.

When  the   Lie group   $SO_\k(2,2)$ is considered, two relevant families of
symmetric  homogeneous spaces~\cite{Gilmore} can be constructed as follows.

\noindent
$\bullet$  The  (2+1)D spacetime.  This is a rank 1 space, associated with the automorphism
$\Pi\Theta$ (\ref{comp}) and Cartan decomposition (\ref{spacea}), which is defined through the quotient 
$$
\mathbb S_{(1)}=SO_\k(2,2)/SO(2,1) ,
$$
  where
$SO(2,1)$ is the Lorentz subgroup spanned by $J$ and $\>K$. Thus
momenta
$P_\mu$ characterize the tangent space at the origin. This space turns out
to have constant curvature equal to the contraction parameter:   $\k=\pm 1/R^2$ for
(A)dS and $\k=0$ ($R\to
\infty$) for Minkowski, where $R$ is the universe radius.

\noindent 
$\bullet$ The 4D space of time-like lines (or wordlines). This is a  rank 2 space, related to the automorphism
$\Pi$ (\ref{bb}) and Cartan decomposition (\ref{spaceb}), which is given by
$$
 \mathbb S_{(2)}=SO_\k(2,2)/\left(SO(2)\otimes
SO_\k(2)\right) ,
$$
where
$SO(2)=\langle J\rangle$ and
$SO_\k(2)=\langle P_0\rangle$.  This space is of hyperbolic type as this has  constant  curvature equal to $-1$ (i.e., $-1/c^2$ in terms of the speed of light). The tangent space is
determined by 
 spatial momenta  $\>P$ and boosts  $\>K$. In fact,  $\mathbb S_{(2)}$ can also be
 interpreted as a $(2\times 2)$D relativistic phase space~\cite{Jaca} in which
position and momentum coordinates are related to the  group
parameters dual to
$\>P$ and $\>K$, respectively. 

We display  in
table \ref{table1} each of the above symmetrical homogeneous spaces for each of the three Lorentzian Lie groups.

\begin{table}[t]
{\footnotesize
 \noindent
\caption{{The AdS, Minkowskian and dS    
 (2+1)D spacetimes and  4D spaces of time-like lines according to the curvature/cosmological constant   $\k=-\Lambda$.}}
\label{table1}
$$
\begin{array}{lll} 
\hline
&&\\[-7pt]
\quad \k ,\ \Lambda&\quad \mbox{Spacetime $\mathbb S_{(1)}$ with curvature $\k$\quad} 
&\quad   \mbox{Space of worldlines $\mathbb S_{(2)}$ with curvature
$-1$\quad}   \\[5pt]
\hline
&&\\[-5pt]
\k>0,\ \Lambda<0  &\quad {\bf AdS}^{2+1}=SO(2,2)/SO(2,1) \quad
&\quad    {\bf LAdS}^{2\times 2}= SO(2,2)/\left(SO(2)\otimes SO(2)\right)
\quad
\\[4pt]
\k= \Lambda=0  &\quad {\bf M}^{2+1}=ISO(2,1)/SO(2,1) 
&\quad    {\bf LM}^{2\times 2}=ISO(2,1)/\left(SO(2)\otimes \R\right) 
\\[4pt]
\k<0,\ \Lambda>0  &\quad {\bf dS}^{2+1}=SO(3,1)/SO(2,1)  
&\quad   {\bf LdS}^{2\times 2}=SO(3,1)/\left(SO(2)\otimes SO(1,1)\right)
\\[4pt]
\hline
\end{array}
$$
}
\end{table}

On the other hand, the two Casimir invariants of the Lie algebra $so_\k(2,2)$ are given by
\be  
{\cal C}=P_0^2-\>P^2+\k(J^2-\>K^2) ,  \qquad
{\cal W}=-JP_0+K_1P_2-K_2P_1  ,
\label{bc} 
\ee  
where $\>P^2=P_1^2+P_2^2$ and $\>K^2=K_1^2+K_2^2$.
Recall that $\cal C$ comes from the Killing--Cartan form, while 
 $\cal W$ is the Pauli--Lubanski vector.


\subsection{Vector model of the (2+1)D relativistic spacetimes}

The action of the (A)dS  groups on  the homogeneous spaces that we have just described is not
linear. As it is well known, this problem can be circumvented by considering the
vector representation of the Lie group $SO_\k(2,2)$ which makes use of an
 ambient space with  an ``extra" dimension. In particular, the  4D real
matrix representation of $so_\k(2,2)$  given by
\be
\begin{array}{l} 
P_0=\left(\begin{array}{cccc}
0&-\k&0&0\cr 
1&0&0&0\cr 
0&0&0&0\cr 
0&0&0&0
\end{array}\right)  ,\quad
P_1=\left(\begin{array}{cccc}
0&0&\k&0\cr 
0&0&0&0\cr 
1&0&0&0\cr 
0&0&0&0
\end{array}\right) , \quad
P_2=\left(\begin{array}{cccc}
0&0&0&\k\cr 
0&0&0&0\cr 
0&0&0&0\cr 
1&0&0&0
\end{array}\right) , \\[25pt]
J=\left(\begin{array}{cccc}
0&0&0&0\cr 
0&0&0&0\cr 
0&0&0&-1\cr 
0&0&1&0
\end{array}\right)  , \quad
K_1=\left(\begin{array}{cccc}
0&0&0&0\cr 
0&0&1&0\cr 
0&1&0&0\cr 
0&0&0&0
\end{array}\right) , \quad
K_2=\left(\begin{array}{cccc}
0&0&0&0\cr 
0&0&0&1\cr 
0&0&0&0\cr 
0&1&0&0
\end{array}\right)  ,
\end{array}
\label{bd}
\ee
fulfils
$$
Y^{T}\mathbb I_{(1)} +\mathbb I_{(1)} Y=0, \qquad Y\in so_\k(2,2) ,\qquad
\mathbb I_{(1)}={\rm diag}\,(1,\k,-\k,-\k) ,
$$
($Y^T$ is the transpose of $Y$).  The exponential of (\ref{bd}) leads
to the vector representation of  
$SO_\k(2,2)$ as a Lie group of  matrices which acts linearly in  
 a 4D space with  ambient (or Weierstrass)
coordinates $(\s_3,\s_\mu)$. The one-parameter subgroups of 
 $SO_\k(2,2)$ obtained from  (\ref{bd}) turn out to be
\be
\begin{array}{l} 
{\rm e}^{x_0 P_0}=\left(\begin{array}{cccc}
\cos \ro x_0&-\ro \sin \ro x_0&0&0\cr 
\frac 1 \ro \sin \ro x_0&\cos \ro x_0&0&0\cr 
0&0&1&0\cr 
0&0&0&1
\end{array}\right) , \quad
{\rm e}^{\theta J}=\left(\begin{array}{cccc}
1&0&0&0\cr 
0&1&0&0\cr 
0&0&\cos\theta&-\sin\theta\cr 
0&0&\sin\theta&\cos\theta
\end{array}\right) ,  \\[25pt]
{\rm e}^{x_1 P_1}=\left(\begin{array}{cccc}
\cosh \ro x_1&0&\ro \sinh \ro x_1&0\cr 
0&1&0&0\cr 
\frac 1 \ro \sinh \ro x_1&0&\cosh \ro x_1&0\cr 
0&0&0&1
\end{array}\right) , \quad
{\rm e}^{\xi_1 K_1}=\left(\begin{array}{cccc}
1&0&0&0\cr 
0&\cosh\xi_1&\sinh\xi_1&0\cr 
0&\sinh\xi_1&\cosh\xi_1&0\cr 
0&0&0&1
\end{array}\right) , \\[2pt]
{\rm e}^{x_2 P_2}=\left(\begin{array}{cccc}
\cosh \ro x_2&0&0&\ro \sinh \ro x_2\cr 
0&1&0&0\cr 
0&0&1&0\cr 
\frac 1 \ro \sinh \ro x_2&0&0&\cosh \ro x_2
\end{array}\right) ,
 \quad
{\rm e}^{\xi_2 K_2}=\left(\begin{array}{cccc}
1&0&0&0\cr 
0&\cosh\xi_2&0&\sinh\xi_2\cr 
0&0&1&0\cr 
0&\sinh\xi_2&0&\cosh\xi_2
\end{array}\right)  ,
\end{array}
\label{dc}
\ee
where hereafter we  also express the curvature as $\k=\ro^2$. Hence,
$\rho=1/R$ and  
$\rho={\rm i}/R$ for  
${\bf AdS}^{2+1}$ and ${\bf dS}^{2+1}$, while the (contraction) limit  $\rho\to
0$ gives rise to
${\bf M}^{2+1}$.

Any element $G\in   SO_\k(2,2)$ verifies that
$G^T\mathbb I_{(1)} G=\mathbb I_{(1)}$.   The (2+1)D spacetime $\mathbb S_{(1)}$ is
identified with the orbit of the origin of the spacetime  
$O=(s_3,s_\mu)=(1,0,0,0)$ which is  contained in the pseudosphere provided by
${\mathbb I}_{(1)}$:
\be
\Sigma_{(1)}\ :\  \s_3^2 +\k (\s_0^2 -  \>\s^2)=1 ,
\label{dd}
\ee
where $\>\s^2=s_1^2+s_2^2$.
The  metric on $\mathbb S_{(1)}$ comes 
from the flat ambient metric  divided by the curvature and
restricted to the above constraint:
\bea
\dd\sigma_{(1)}^2 &\!\!\!=\!\!\!&  \left.{1\over\k}
\left(\dd s_3^2+\k (\dd\s_0^2 -  \dd\s_1^2-\dd\s_2^2)\right)
\right|_{\Sigma_{(1)}} 
\nonumber\\
&\!\!\!=\!\!\!&
\dd\s_0^2 -  \dd\s_1^2-\dd\s_2^2+\k\,\frac{(\s_0 \dd\s_0  - 
\s_1\dd\s_1 -\s_2\dd\s_2 )^2}{1-\k(\s_0^2 -  \>\s^2)}.
\label{de}
\eea

Ambient coordinates can be parametrized in terms of three  intrinsic
spacetime coordinates in different ways. We shall introduce the geodesic
parallel coordinates $x_\mu$ \cite{conformes} through
the following action of the momenta subgroups (\ref{dc}) on the origin:
$$
(\s_3,\s_\mu )(x_\nu)=\exp(x_0 P_0)\exp(x_1 P_1)\exp(x_2 P_2)  O ,
$$
namely,
\be
\begin{array}{l} 
\s_3=\cos \ro x_0 \cosh \ro x_1 \cosh \ro x_2 ,\\ 
\displaystyle {\s_0=\frac {\sin \ro x_0}\ro  \cosh \ro x_1 \cosh \ro
x_2 }, \\[8pt]  
\displaystyle {\s_1=\frac {\sinh \ro x_1 }\ro   \cosh \ro x_2 } ,\\[8pt] 
\displaystyle {\s_2=\frac { \sinh \ro x_2}\ro  . }
\end{array}
\label{dg}
\ee
The role of the coordinates $x_\mu$ that parametrize a generic point $Q$ under
(\ref{dg}) in the   (2+1)D spacetime  is the following. Let $l_0$ a
time-like geodesic and $l_1$, $l_2$ two space-like geodesics such that
these three basic geodesics are orthogonal at
$O$. Then   $x_0$ is the geodesic distance from $O$ up to a point
$Q_1$ measured along  $l_0$;  $x_1$ is the distance between $Q_1$ and another
point $Q_2$ along a space-like geodesic $l'_1$ orthogonal to $l_0$
through $Q_1$ and parallel to $l_1$; and  
$x_2$ is the distance between $Q_2$ and  $Q$ 
along a space-like geodesic $l'_2$ orthogonal
to $l'_1$ and parallel to $l_2$. 
Recall that time-like geodesics (as $l_0$) are compact in ${\bf
AdS}^{2+1}$ and non-compact in
${\bf dS}^{2+1}$,  while space-like ones (as  $l_i$, $l'_i$) are
compact in
${\bf dS}^{2+1}$ but non-compact in  ${\bf AdS}^{2+1}$. Thus  the
trigonometric functions depending on $x_0$ are circular in  ${\bf
AdS}^{2+1}$ ($\rho=1/R)$ and hyperbolic in ${\bf dS}^{2+1}$ 
($\rho={\rm i}/R)$  and, conversely, those depending on $x_i$ are
circular in  ${\bf dS}^{2+1}$  but hyperbolic in ${\bf AdS}^{2+1}$.

Under (\ref{dg}), the   metric (\ref{de}) now reads
\be
\dd\sigma_{(1)}^2 =\cosh^2(\ro x_1) \cosh^2(\ro x_2)\dd x_0^2-\cosh^2(\ro
x_2)\dd x_1^2-\dd x_2^2 .
\label{dh}
\ee
Notice that if $\ro\to 0$, the
parametrization (\ref{dg}) gives the  flat Cartesian coordinates 
$\s_3=1,  \s_\mu= x_\mu$, and the  metric (\ref{dh}) reduces to 
$\dd\sigma_{(1)}^2 = \dd x_0^2- \dd
x_1^2-\dd x_2^2$ in ${\bf M}^{2+1}$.


\subsection{Bivector model of the 4D   spaces of  worldlines}

The action of $SO_\k(2,2)$ on the space  of time-like lines $\mathbb S_{(2)}$ is
also a non-linear one. As in the previous case, this problem can be solved by
introducing an ambient space, now  6D with two ``extra"
dimensions, on which the group acts linearly and where  $\mathbb S_{(2)}$ is
embedded. Let us consider the so called bivector representation of (\ref{ba})   
given by
\cite{Jaca}:
\be
\begin{array}{ll}  
P_0=-\k e_{24}+e_{42}-\k e_{35}+e_{53} , \quad &
J=-e_{23}+e_{32}-e_{45}+e_{54} , \\ 
P_1=-\k e_{14}-e_{41}+\k
e_{36}+e_{63} ,\quad &  K_1=e_{12}+e_{21}+e_{56}+e_{65}  ,
\\ 
P_2=-\k e_{15}-e_{51}-\k e_{26}-e_{62}, \quad
&K_2=e_{13}+e_{31}-e_{46}-e_{64}     ,
\end{array}
\label{di}
\ee
where $e_{ab}$ is the $6\times 6$ matrix with entries
$(e_{ab})_{ij}=\delta_{ai}\delta_{bj}$. Under this representation   any generator $Y\in
so_\k(2,2)$   fulfils
$$
Y^{T}  \mathbb I_{(2)}  +\mathbb I_{(2)}  Y=0  , \qquad \mathbb I_{(2)} ={\rm
diag}\,(1,-1,-1,-\k,-\k,\k).
$$
By exponentiation of (\ref{di}) we obtain the bivector representation of
$SO_\k(2,2)$, that is,     a group of matrices which acts linearly in a 6D
 space with ambient (or Pl\"ucker) coordinates 
 $(\et_3,\et_1,\et_2,\yy_1,\yy_2,\yy_3)$. The origin   of $\mathbb
S_{(2)}$ is ${\cal O}=(1,0,0,0,0,0)$ and this space   is identified with the
intersection of the pseudosphere
$\Sigma_{(2)}$ determined by $\mathbb I_{(2)}$ with a quadratic cone
$\mathbb P$ known as Pl\"ucker or Grassmann relation (invariant under the
group action); these constraints are given by \cite{Jaca}:
\be
\Sigma_{(2)}\ :\ \et_3^2-\boldsymbol{\et}^2+\k(\yy^2_3-\boldsymbol{\yy}^2)=1 ,
 \qquad  
\mathbb P\ :\  \et_3 \yy_3-\et_1 \yy_2+\et_2\yy_1=0,
\label{dk}
\ee
where $\boldsymbol{\et}^2=\et_1^2+ \et_2^2$ and $\boldsymbol{\yy}^2=\yy_1^2+ \yy_2^2$.
The metric on $\mathbb S_{(2)}$ follows from the 6D flat ambient metric 
divided by the negative curvature of  $\mathbb S_{(2)}$  and  subjected
to both conditions (\ref{dk}):
 \be
\dd\sigma_{(2)}^2  =   \left.{1\over{-1}}
\left(
\dd\et_3^2-\dd\et_1^2-\dd\et_2^2+\k(\dd\yy^2_3-\dd\yy_1^2-\dd\yy_2^2)
\right)
\right|_{\Sigma_{(2)}, \mathbb P} .
\label{dl}
\ee

Pl\"ucker coordinates can   be expressed  through four intrinsic
quantities of $\mathbb S_{(2)}$. We shall consider the space $\>x$ and
rapidity (boost) $\boldsymbol{\xi}$ group coordinates. The action of the following
sequence of one-parameter subgroups on $\cal O$ (those defining the tangent
space to  $\mathbb S_{(2)}$) under the representation  (\ref{di}),
$$
(\et_3,\boldsymbol{\et},\boldsymbol{\yy},
\yy_3)(\>x,\boldsymbol{\xi})= \exp(x_1 P_1)\exp(x_2
P_2)\exp(\xi_1 K_1)\exp(\xi_2 K_2){\cal O}  ,
$$
gives rise to 
\be
\begin{array}{l} 
\et_3= \cosh \ro x_1\cosh \ro x_2\cosh  \xi_1\cosh \xi_2 , \\[2pt]
\et_1=  \cosh \ro x_2\sinh  \xi_1\cosh \xi_2 ,\\[2pt]
\et_2=\cosh \ro x_1 \sinh \xi_2 - \sinh \ro x_1\sinh \ro x_2\sinh 
\xi_1\cosh \xi_2 ,\\[2pt]
\displaystyle{ \yy_1= -\frac{\sinh \ro x_1}{\ro}\cosh \ro x_2\cosh 
\xi_1\cosh
\xi_2} ,\\[8pt]
\displaystyle{  \yy_2=- \frac{\sinh \ro x_2}{\ro}\cosh  \xi_1\cosh \xi_2
}, \\[8pt]
\displaystyle{  \yy_3 =  \frac{\sinh \ro x_1}{\ro} \sinh \xi_2 -
 \cosh \ro x_1\frac{\sinh \ro x_2}{\ro}\sinh 
\xi_1\cosh \xi_2   } .
\end{array}
\label{dn}
\ee

Under the contraction $\rho\to 0$  ($\k=0$),   this parametrization reduces
to that of the 4D Minkows\-kian space of worldlines    $ {\bf LM}^{2\times 2}$: 
\be
\begin{array}{ll} 
\et_3= \cosh  \xi_1\cosh \xi_2 ,&\quad  
\displaystyle{  \yy_3 =   x_1  \sinh \xi_2 -
    x_2 \sinh  \xi_1\cosh \xi_2   }  , \\[1pt]
\et_1=   \sinh  \xi_1\cosh \xi_2 ,&\quad \displaystyle{ \yy_1= -    x_1  \cosh 
\xi_1\cosh \xi_2} ,\\[2pt]
\et_2= \sinh \xi_2   , &\quad \displaystyle{  \yy_2=-    x_2 \cosh  \xi_1\cosh
\xi_2 } . 
\end{array}
\label{do}
\ee
Such expressions indicate  that   the Pl\"ucker coordinates
 $(\boldsymbol{\yy}, \yy_3)$  and  $(\boldsymbol{\et}, \et_3)$ can be
interpreted as ``position-like"  and ``momentum-like"   ones, respectively, within
the phase space $(\>x,\boldsymbol{\xi})$~\cite{Jaca}.  In $ {\bf
LM}^{2\times 2}$  the metric (\ref{dl}) is degenerate and reads
$$
\dd\sigma_{(2)}^2  =   
 \dd\et_1^2+\dd\et_2^2-\frac{(\et_1 
\dd\et_1+\et_2\dd\et_2
)^2}{1+\boldsymbol{\et}^2}=\cosh^2\xi_2\,\dd\xi_1^2+\dd\xi_2^2 ,
$$
which corresponds to a 2D space of rank 1 with
negative curvature, that is,   the 2D velocity 
Minkowskian space (so with coordinates $\boldsymbol{\xi}$) is hyperbolic.  Nevertheless, we stress
that this is no longer true when
$\k\ne 0$ (in both  $ {\bf LAdS}^{2\times 2}$ and 
$ {\bf LdS}^{2\times 2}$) where the complete $(2\times 2)$D space
structure is required (so with the four  coordinates $(\>x,\boldsymbol{\xi})$), thus precluding the possibility of using   a ``reduced" 2D
velocity space.


\section{(Anti-)de Sitter Drinfel'd-doubles and first-order\\ non-commutative  spaces}

The first-order deformation terms in the coproduct of the   $\kappa$-Poincar\'e algebra
\cite{LukierskiRuegg1992,Giller,Lukierskib,Maslanka,Majid:1994cy,Zak,LukR,LukNR}
are known to be generated  by the following  classical $r$-matrix:
\be
r=z(K_1\wedge P_1+K_2\wedge P_2) ,
\label{ca}
\ee
where $\wedge$ denotes the skewsymmetric tensor product. Recall that $r$ is a solution of the modified classical Yang--Baxter equation and that
$z$ is related to the
usual $\kappa$ and $q$ deformation  parameters by $z=1/\kappa=\ln q$.

Such a classical
$r$-matrix also holds for the (A)dS  algebras \cite{LBC}, so that we
shall consider (\ref{ca}) for the  whole family $so_\k(2,2)$. 
Hence this element
gives rise to the cocommutator $\delta$ of any generator $Y_i$ through
the relation  $\delta(Y_i)=[Y_i\otimes 1+1\otimes Y_i,r]$, namely
\be 
\begin{array}{l} 
\delta(P_0)=0  , \qquad \delta(J)=0 , \\[2pt]
\delta(P_i)=z(P_i\wedge P_0-\k\epsilon_{ij} K_j\wedge J),  \\[2pt]
\delta(K_i)=z(K_i\wedge P_0+\epsilon_{ij} P_j\wedge J) . 
\end{array}
\label{cb}
\ee 
Next if we denote by $\hat y^i$ the    quantum group coordinate dual to
$Y_i$,  such that  $\langle \hat y^i| Y_j\rangle=    \delta_j^i$, and
write the cocommutators as  $\delta(Y_i)=f_i^{jk}Y_j\wedge Y_k$, then 
   Lie bialgebra     duality  provides the 
so called Drinfel'd-double Lie algebra~\cite{CP,Majid}  formed by three sets of
brackets: the initial Lie algebra, the dual relations and the crossed
commutation rules, namely
$$
[Y_i,Y_j]=c_{ij}^k Y_k , \qquad 
[\hat y^i,\hat y^j]=f_k^{ij}\hat y^k  ,\qquad
[\hat y^i,Y_j]=c_{jk}^i \hat y^k -  f_j^{ik} Y_k .
$$
The cocycle condition for the cocommutator $\delta$   implies the following compatibility equations among the structure constants $c_{ij}^k$ and $f_k^{ij}$:
$$
f^{ab}_k c^k_{ij} = f^{ak}_i c^b_{kj}+f^{kb}_i c^a_{kj}
+f^{ak}_j c^b_{ik} +f^{kb}_j c^a_{ik}. 
$$

In our case,   we denote by   $\{ \hat\theta, \hat
x_\mu,  \hat\xi_i \} $ the dual non-commutative coordinates of the 
generators $\{J,P_\mu,K_i\}$, respectively. Thus the  (A)dS and
Poincar\'e Drinfel'd-doubles   are collectively given in terms
of the curvature
$\k$ and deformation parameter $z$   by the initial Lie algebra
${so}_\k(2,2)$ (\ref{ba}), the dual commutators
\be 
\begin{array}{llll} 
[\hat\theta,\hat x_i]= z\epsilon_{ij}  \hat \xi_j , &\quad
[\hat x_0, \hat x_i]=-z \hat x_i  ,&\quad
[\hat x_1, \hat x_2]=0,&\quad [\hat\theta,\hat x_0]=0, \\[2pt]
[\hat\theta,\hat \xi_i]=-z\k \epsilon_{ij}  \hat x_j ,&\quad
[\hat x_0, \hat \xi_i]=-z \hat \xi_i ,&\quad
[\hat \xi_1, \hat \xi_2]=0,&\quad [\hat x_i,\hat \xi_j]=0 ,
\end{array}
\label{cd}
\ee 
together with the crossed relations
$$
\begin{array}{ll} 
[\hat x_0,J]=[\hat x_0,P_0]=0 ,&\qquad
[\hat\theta,J]=[\hat \theta,P_0]=0 ,\\[2pt]
[\hat x_0,P_i]=-(\hat\xi_i-zP_i) ,&\qquad
[\hat\theta,P_i]=-\k\epsilon_{ij}(\hat x_j+z K_j), \\[2pt]
[\hat x_0,K_i]= \hat x_i+z K_i ,&\qquad
[\hat \theta,K_i]= - \epsilon_{ij} (\hat\xi_j-z P_j), \\[2pt]
[\hat x_i,J]=- \epsilon_{ij} \hat x_j  ,&\qquad
[\hat \xi_i,J]=- \epsilon_{ij} \hat\xi_j , \\[2pt]
[\hat x_i,P_0]=-\hat\xi_i ,&\qquad  [\hat \xi_i,P_0]=\k\hat x_i ,\\[2pt]
[\hat x_i,P_j]=\epsilon_{ij} \hat\theta -z \delta_{ij} P_0, &\qquad
[\hat \xi_i,P_j]=-\k(\delta_{ij} \hat x_0+z\epsilon_{ij} J) ,\\[2pt]
[\hat x_i,K_j]=\delta_{ij} \hat x_0+z\epsilon_{ij} J ,&\qquad
[\hat \xi_i,K_j]=\epsilon_{ij} \hat\theta -z \delta_{ij} P_0 .
\end{array}
$$

Parity and time-reversal automorphisms (\ref{bb}) can be generalized to the full  
Drinfel'd-double   as follows
\be 
\begin{array}{ll} 
\Pi_z  :& (P_0, \>P , \>K ,J;\ \hat x_0,\hat \>x  ,\hat{ \boldsymbol{ \xi}}, \hat
\theta ;\ z)\to  (P_0,- \>P ,- \>K ,J;\  \hat x_0,-\hat \>x ,-\hat{ \boldsymbol{
\xi}},
\hat
\theta;\ z)  ,\\[2pt]
\Theta_z  :& (P_0, \>P , \>K ,J;\  \hat x_0,\hat \>x,\hat{ \boldsymbol{ \xi}},
\hat
\theta ;\ z)\to  (-P_0, \>P ,- \>K ,J;\ -\hat x_0,\hat{ \>x},-\hat{
\boldsymbol{
\xi}},
\hat
\theta;\ -z). 
\end{array}
\label{cf}
\ee

Since the first-order structure  of the complete quantum deformation of
${so}_\k(2,2)$ is described by the corresponding Drinfel'd-double,
some preliminary information concerning the physical properties  of  the
associated non-commutative spaces can   be  extracted from it.   Notice that, in this
first-order approach, all the expressions will be linear both on the generators and
on the dual quantum group coordinates.


\subsection{Non-commutative  spacetimes: linear relations}

The usual way to propose a non-commutative spacetime is to consider the
commutation rules involving the quantum coordinates $\hat x_\mu$.
Therefore, from (\ref{cd}) we find that the three (A)dS and Minkowskian 
non-commutative spacetimes are simultaneously defined by the {\em same}   first-order
relations:
\be
[\hat x_0, \hat x_i]=-z \hat x_i ,  \qquad
[\hat x_1, \hat x_2]=0 ,
\label{ea}
\ee
which coincide with   the $\kappa$-Minkowski
space,  ${\bf M}_z^{2+1}$,~\cite{Maslanka,Majid:1994cy,Zak,LukR} for any
 value of the curvature $\k$. As we shall see in section 5, further corrections of
(\ref{ea}) depending on $\k$ will appear when the full quantum (A)dS groups
are considered.

As it was already studied in~\cite{Kowalskia,Kowalskib}, the action of the
generators on the non-commuta\-tive   spacetime follows by replacing formally
$P_\mu\to
\hat x_\mu$, which requires to consider the commutators involving $\{J,\hat
x_\mu,K_i\}$ within the Drinfel'd-double. Next the   change of basis given
by~\cite{Kowalskib}:
$$
\qq_0=-\hat x_0  , \qquad
\qq_i=\hat x_i+z K_i  ,
\label{eb} 
$$
provides the following commutation relations
 \be 
\begin{array}{lll} 
[J,\qq_i]=   \epsilon_{ij}\qq_j  ,&\qquad
[J,K_i]=   \epsilon_{ij}K_j , &\qquad  [J,\qq_0]= 0 ,  \\[2pt]
[\qq_i,K_j]=-\delta_{ij}\qq_0, &\qquad [\qq_0,K_i]=-\qq_i ,&\qquad
[K_1,K_2]= -J   , \\[2pt]
[\qq_0,\qq_i]=-z^2 K_i,  &\qquad [\qq_1,\qq_2]= z^2 J  ,
\end{array}
\label{ec}
\ee 
that can directly be  related to the   initial Lie algebra (\ref{ba}). Consequently,
whenever $z$ is a {\em real} deformation parameter, the commutators
(\ref{ec}), that do not depend on $\k$, close the  Lie algebra   $so(3,1)$   for
the three cases. Hence we obtain the dS spacetime as  the homogeneous space 
$$
{\bf dS}^{2+1}\equiv  \langle J,\>K, \qq_\mu\rangle/\langle J,\>K \rangle = SO(3,1)/SO(2,1),
\nonumber
$$
such that the deformation parameter now plays the role of the (negative) curvature equal to  $-z^2$.

 We
stress that the connection between  ${\bf M}_z^{2+1}$ and the dS space was so
established in~\cite{Kowalskia} and further developed in~\cite{Kowalskib}, so that the
expressions (\ref{ec})   generalize such a link for the non-commutative (A)dS
cases as well.


\subsection{Non-commutative   spaces of worldlines: linear relations}

A similar procedure suggests that the corresponding non-commutative spaces of
worldlines  arise within the Drinfel'd-double through the
commutators of $\hat \>x$ and $\hat{ \boldsymbol{ \xi}}$ (dual to
$\>P$ and $\>K$); these are  
\be
[\hat x_1, \hat x_2]=0  , \qquad
[\hat \xi_1, \hat \xi_2]=0  ,\qquad [\hat x_i,\hat \xi_j]=0   ,
\label{ed}
\ee
which are  trivially  independent of $z$ and $\k$.  

 The adjoint action on the quantum coordinates $\hat \>x$,  $\hat{ \boldsymbol{
\xi}}$ of the isotropy subgroup of a worldline  spanned by
$J$ and $P_0$ (\ref{spaceb})
gives the following non-deformed commutation rules:
 \be 
\begin{array}{lll} 
[J,\hat x_i]=   \epsilon_{ij}\hat x_j  ,&\qquad
[J,\hat \xi_i]=   \epsilon_{ij}\hat \xi_j , &\qquad  [J,P_0]= 0 ,  \\[2pt]
[\hat x_i,\hat \xi_j]=0 ,&\qquad [P_0,\hat \xi_i]=-\k \hat x_i ,&\qquad
[\hat \xi_1, \hat \xi_2]=0  , \\[2pt]
[\hat x_1, \hat x_2]=0 ,&\qquad [P_0,\hat x_i]= \hat \xi_i ,
\end{array}
\label{ef}
\ee 
where the deformation parameter $z$ does not appear.
To unveil this structure we rename the former generators   as 
$$
J'=J,\qquad P'_0=-P_0 ,\qquad P'_i=\hat \xi_i ,\qquad K'_i=\hat x_i ,
$$
and 
 from   (\ref{ef}) we
obtain the commutation relations
  \be  
\begin{array}{lll} 
[J',P'_i]=   \epsilon_{ij}P'_j  ,&\qquad
[J',K'_i]=   \epsilon_{ij}K'_j  ,&\qquad  [J',P'_0]= 0 ,  \\[2pt]
[P'_i,K'_j]=0 ,&\qquad [P'_0,K'_i]=-P'_i ,&\qquad
[K'_1,K'_2]=0  , \\[2pt]
[P'_0,P'_i]=\k K'_i ,&\qquad [P'_1,P'_2]=0  .
\end{array}
\label{babis}
\ee 
Surprisingly enough, these relations define just the  Newtonian Lie algebras coming from    the non-relativistic limit $c\to \infty$ of the three Lorentzian Lie algebras (\ref{ba}), keeping the constant curvature $\k$. Namely, the relations (\ref{babis})  close  the 
oscillating  Newton--Hooke, Galilei
and expanding  Newton--Hooke algebras~\cite{CK3,Ober2,BLL,Aldrovandi} according to  
$\k>,=,<0$,  
respectively.  This fact is consistent with the   known result that
establishes that  each of the    non-relativistic 
Newtonian spaces of constant curvature $\k$ can be obtained  from  the
corresponding  relativistic   one  through a contraction around a time-like line. Therefore  the classical (non-deformed) picture is preserved for
worldlines.

Summing up, 
 the  first-order
deformation of $so_\k(2,2)$ characterized by the chosen $r$-matrix (\ref{ca})
conveys   non-commutativity on the spacetime (\ref{ea}) but commutativity on the space of
worldlines (\ref{ed}). Furthermore,  space isotropy  is ensured in both cases  as the corresponding commutation relations
  do not involve the quantum rotation
coordinate $\hat \theta$.


\section{A Poisson--Lie structure  on the (anti-)de Sitter 
groups}

So far we have studied the first-order quantum (A)dS  deformation.
However, the obtention of   the complete  (in all
orders in $z$ and in the generators)       deformation  of a semisimple group   in terms of local coordinates
is, in general,  a very involved task.
A way to study the non-commutative structures   is to  compute the  Poisson--Lie
brackets (derived from (\ref{ca})) for the commutative coordinates and next to
analysing their possible non-commutative version.

In particular, let us consider the  $4\times 4$ matrix element of the group
$SO_\k(2,2)$ obtained through the following product written under the
representation (\ref{dc}):
\be
T=\exp(x_0 P_0)\exp(x_1 P_1)\exp(x_2 P_2) \exp(\xi_1 K_1)\exp(\xi_2 K_2)
\exp(\theta J)  ,
\label{ga}
\ee
where the group coordinates are commutative ones. Left-  and
right-invariant vector fields, $Y^L$ and $Y^R$, of $SO_\k(2,2)$
deduced from (\ref{ga}) are displayed in table~\ref{table2}. Notice that such expressions hold for any value of the curvature $\k$. In the Poincar\'e case with $\k=0$, the vector fields,  coming  from the    smooth contraction limit $\rho\to 0$,  
are rather simplified  as shown in table~\ref{table3}.

\begin{table}[htbp]
{\footnotesize
 \noindent
\caption{{Left- and right-invariant (anti-)de Sitter vector fields with $\k=\rho^2=-\Lambda$.}}
\label{table2}
$$
\begin{array}{l} 
\hline
 \\[-7pt]
\multicolumn1{c}{  \mbox{Left-invariant vector fields} }  \\[4pt]
\hline
 \\[-5pt]
\displaystyle{  Y_{P_0}^L=\frac{\cosh\xi_1\cosh\xi_2}{\cosh \rho x_1\cosh \rho
x_2}\,( \partial_{x_0} -\rho \sinh \rho x_1\partial_{\xi_1})
+\frac{\sinh\xi_1\cosh\xi_2}{ \cosh
\rho x_2}\,\partial_{x_1}+\sinh\xi_2\,\partial_{x_2}  -\rho \cosh\xi_2\tanh \rho
x_2\,\partial_{\xi_2}  } \\[12pt]
\displaystyle{   Y_{P_1}^L=\left(
\frac{\cos\theta \sinh\xi_1+\sin\theta\cosh\xi_1\sinh\xi_2}{\cosh \rho
x_1\cosh
\rho x_2}\right) \! ( \partial_{x_0} -\rho \sinh \rho x_1\partial_{\xi_1}) +\left(
\frac{\cos\theta \cosh\xi_1+\sin\theta\sinh\xi_1\sinh\xi_2}{ \cosh
\rho x_2}\right)  \! \partial_{x_1}   }\\[8pt]
\displaystyle{\qquad\qquad +\sin\theta\cosh \xi_2\,\partial_{x_2}
-\rho  \cos\theta \tanh \rho x_2 \tanh\xi_2 
\,\partial_{\xi_1}-\rho  \sin\theta \tanh \rho x_2 \sinh\xi_2 
\,\partial_{\xi_2}+ \rho \, \frac{\cos\theta   \tanh \rho x_2 }{\cosh\xi_2 } 
\,\partial_{\theta}  }\\[12pt]
\displaystyle{   Y_{P_2}^L=\left(
\frac{\cos\theta \cosh\xi_1\sinh\xi_2-\sin\theta \sinh\xi_1}{\cosh \rho
x_1\cosh
\rho x_2}\right)  \!  ( \partial_{x_0} -\rho \sinh \rho x_1\partial_{\xi_1}) +\left(
\frac{\cos\theta\sinh\xi_1\sinh\xi_2-\sin\theta \cosh\xi_1}{ \cosh
\rho x_2}\right)  \!  \partial_{x_1}  }\\[8pt]
\displaystyle{\qquad\qquad +\cos\theta\cosh \xi_2\,\partial_{x_2}
+\rho  \sin\theta \tanh \rho x_2 \tanh\xi_2
\,\partial_{\xi_1}-\rho  \cos\theta\sinh\xi_2 \tanh \rho x_2 
\,\partial_{\xi_2}- \rho   \,\frac{\sin\theta\tanh \rho x_2 }{\cosh\xi_2 } 
\,\partial_{\theta}  }\\[12pt]
\displaystyle{   Y_{K_1}^L= 
\frac{\cos\theta  }{\cosh  
\xi_2} \,\partial_{\xi_1} +\sin\theta\,\partial_{\xi_2}+\cos\theta  \tanh  
\xi_2\,\partial_{\theta}  }\\[12pt]
\displaystyle{   Y_{K_2}^L= -
\frac{\sin\theta  }{\cosh  
\xi_2} \,\partial_{\xi_1} +\cos\theta\,\partial_{\xi_2}-\sin\theta  \tanh  
\xi_2\,\partial_{\theta}  }\\[12pt]
\displaystyle{     Y_{J}^L= \partial_{\theta} }\\[6pt]
 \hline
 \\[-7pt]
\multicolumn1{c}{  \mbox{Right-invariant vector fields} }  \\[4pt]
\hline
 \\[-5pt]
\displaystyle{  Y_{P_0}^R= \partial_{x_0}   } \\[10pt]
\displaystyle{   Y_{P_1}^R= 
-\sin \rho x_0\tanh \rho x_1 \, \partial_{x_0} 
+\cos \rho x_0\, \partial_{x_1} -\rho\,\frac{\sin \rho x_0}{\cosh \rho
x_1}\, \partial_{\xi_1}  }\\[12pt] 
\displaystyle{   Y_{P_2}^R=
-\frac{\sin \rho x_0\tanh \rho x_2 }{\cosh \rho x_1} \,\partial_{x_0} 
-\cos \rho x_0\sinh \rho x_1\tanh \rho x_2\,\partial_{x_1}+
\cos \rho x_0\cosh \rho x_1\,\partial_{x_2}  }\\[8pt]
\displaystyle{\qquad\qquad + \rho \sin \rho x_0\tanh \rho x_1\tanh \rho x_2\,
 \partial_{\xi_1}+  {\rho} \left( \frac{ \cos \rho x_0\sinh \rho
x_1\sinh  \xi_1- \sin \rho x_0 \cosh  \xi_1}{\cosh \rho x_2} \right) 
 \partial_{\xi_2}  }\\[8pt]
\displaystyle{\qquad\qquad +   {\rho}\left( \frac
{\cos \rho x_0\sinh \rho x_1\cosh  \xi_1- \sin \rho x_0 \sinh  \xi_1} {\cosh \rho x_2
\cosh  \xi_2}\right) 
\left( \partial_{\theta} -  \sinh\xi_2 \partial_{\xi_1} \right)  }\\[12pt]
\displaystyle{   Y_{K_1}^R= 
\frac{\cos \rho x_0  \tanh \rho x_1}{\rho}\,\partial_{x_0} +\frac{\sin \rho
x_0}{\rho}\,\partial_{x_1} +\frac{\cos \rho x_0}{\cosh \rho x_1}\,
\partial_{\xi_1}   }\\[12pt]
\displaystyle{    Y_{K_2}^R= 
\frac{\cos \rho x_0  \tanh \rho x_2}{\rho \cosh \rho x_1}\,\partial_{x_0}
-\frac{\sin \rho x_0  \sinh \rho x_1  \tanh \rho x_2}{\rho}\,\partial_{x_1}
+\frac{\sin \rho x_0 \cosh \rho x_1}{\rho}\,\partial_{x_2}   }
\\[8pt]
\displaystyle{\qquad\qquad -  \cos \rho x_0\tanh \rho x_1\tanh \rho x_2\,
 \partial_{\xi_1}+ \left( \frac{ \sin \rho x_0  \sinh\rho  x_1
\sinh  \xi_1  + \cos \rho x_0\cosh  
\xi_1 }{\cosh \rho x_2}  \right)  \partial_{\xi_2}  }\\[8pt]
\displaystyle{\qquad\qquad +   \left( \frac
{\sin \rho x_0\sinh \rho x_1\cosh  \xi_1+ \cos \rho x_0 \sinh  \xi_1} { \cosh \rho x_2
\cosh  \xi_2  }  \right) 
\left( \partial_{\theta} - \sinh\xi_2 \partial_{\xi_1}
\right)  }\\[12pt]
\displaystyle{    Y_{J}^R= 
-\frac{\cosh \rho x_1  \tanh \rho x_2}{\rho  }\,\partial_{x_1}
+\frac{  \sinh \rho x_1 }{\rho}\,\partial_{x_2}
- \frac{\cosh \rho x_1 }{\cosh \rho
x_2}\left(\cosh\xi_1 \tanh\xi_2 \, \partial_{\xi_1}  - \sinh\xi_1  \,
\partial_{\xi_2}  -  \frac{\cosh\xi_1  }{\cosh\xi_2 }\,\partial_{\theta} \right)
 }\\[10pt]
\hline
\end{array}
$$
}
\end{table}

\begin{table}[htbp]
{\footnotesize
 \noindent
\caption{{Left- and right-invariant Poincar\'e vector fields with $\k=\rho=\Lambda=0$.}}
\label{table3}
$$
\begin{array}{l} 
\hline
 \\[-7pt]
\multicolumn1{c}{  \mbox{Left-invariant vector fields} }  \\[4pt]
\hline
 \\[-5pt]
\displaystyle{  Y_{P_0}^L= {\cosh\xi_1\cosh\xi_2} \,  \partial_{x_0}  
+ {\sinh\xi_1\cosh\xi_2} \,\partial_{x_1}+\sinh\xi_2\,\partial_{x_2}  } \\[8pt]
\displaystyle{   Y_{P_1}^L=\left(
 {\cos\theta \sinh\xi_1+\sin\theta\cosh\xi_1\sinh\xi_2} \right)  \partial_{x_0}  +\left(
 {\cos\theta \cosh\xi_1+\sin\theta\sinh\xi_1\sinh\xi_2} \right)\partial_{x_1} +\sin\theta\cosh \xi_2\,\partial_{x_2}  }\\[8pt]
\displaystyle{   Y_{P_2}^L=\left(
 {\cos\theta \cosh\xi_1\sinh\xi_2-\sin\theta \sinh\xi_1} \right)  \partial_{x_0} +\left(
 {\cos\theta\sinh\xi_1\sinh\xi_2-\sin\theta \cosh\xi_1} \right)\partial_{x_1}  +\cos\theta\cosh \xi_2\,\partial_{x_2} }\\[8pt]
\displaystyle{   Y_{K_1}^L= 
\frac{\cos\theta  }{\cosh  
\xi_2} \,\partial_{\xi_1} +\sin\theta\,\partial_{\xi_2}+\cos\theta  \tanh  
\xi_2\,\partial_{\theta}  }\\[10pt]
\displaystyle{   Y_{K_2}^L= -
\frac{\sin\theta  }{\cosh  
\xi_2} \,\partial_{\xi_1} +\cos\theta\,\partial_{\xi_2}-\sin\theta  \tanh  
\xi_2\,\partial_{\theta}  }\\[10pt]
\displaystyle{     Y_{J}^L= \partial_{\theta} }\\[6pt]
 \hline
 \\[-7pt]
\multicolumn1{c}{  \mbox{Right-invariant vector fields} }  \\[4pt]
\hline
 \\[-5pt]
\displaystyle{  Y_{P_0}^R= \partial_{x_0}  \qquad Y_{P_1}^R= \partial_{x_1}  \qquad Y_{P_2}^R= \partial_{x_2}  } \\[8pt]
\displaystyle{   Y_{K_1}^R= 
 x_1\, \partial_{x_0} + x_0\,\partial_{x_1} + \partial_{\xi_1}   }\\[8pt]
\displaystyle{    Y_{K_2}^R= 
x_2\,\partial_{x_0}
+x_0\,\partial_{x_2} -    {\sinh\xi_1} {\tanh\xi_2} \, \partial_{\xi_1} + \cosh  
\xi_1\,\partial_{\xi_2}   +    \frac{\sinh\xi_1}{\cosh\xi_2} \,\partial_{\theta}}
\\[10pt]
\displaystyle{    Y_{J}^R=  - x_2\,\partial_{x_1} + x_1\,\partial_{x_2}
-  \cosh\xi_1 \tanh\xi_2 \, \partial_{\xi_1} + \sinh\xi_1  \,
\partial_{\xi_2}  +  \frac{\cosh\xi_1  }{\cosh\xi_2 }\,\partial_{\theta}  
 }\\[10pt]
\hline
\end{array}
$$
}
\end{table}

The Poisson--Lie brackets that close
the algebra of smooth functions on  the (A)dS groups, ${\rm Fun}(SO_\k(2,2))$ (r.h.s.\ of the diagram (\ref{aa})),
associated to an  
$r$-matrix   $r= 
r^{ij}Y_i\otimes Y_j$  come  from the Sklyanin bracket defined by
\cite{Drlb}: 
\be
\{f,g\}=  r^{ij}(Y_i^Lf\, Y_j^L g -
Y_i^Rf\, Y_j^R g)  ,\qquad f,g\in   {\rm Fun}(SO_\k(2,2)).
\label{gb}
\ee
Thus by substituting the vector fields of table \ref{table2} and   the
classical $r$-matrix (\ref{ca}) in (\ref{gb}) we obtain the 
Poisson--Lie brackets between the six  {\em commutative} group coordinates  $\{ 
\theta,   x_\mu,  \xi_i \} $  which
are splitted  in  the following three sets.

\noindent
$\bullet$ Those involving spacetime  $x_\mu$ group
coordinates:
  \be  
\begin{array}{l} 
\displaystyle{ \{x_0,x_1\} =-z\,\frac{\tanh\rho x_1}{\rho \cosh^2\!\rho x_2} , \qquad
\{x_0,x_2\} =-z\,\frac{\tanh\rho x_2}{\rho } ,\qquad 
\{x_1,x_2\} =0 . } 
\end{array}
\label{gc}
\ee 
 
\noindent
$\bullet$ Those that comprise  space  $\>x$ and boost ${ \boldsymbol{
\xi}}$ coordinates (besides the above vanishing bracket):
\be  
\begin{array}{l} 
\displaystyle{ \{x_1,\xi_1\} = \frac{z}{  \cosh \rho x_2}\left(
\frac{\cosh \rho x_2}{\cosh \rho x_1}-\frac{\cosh \xi_1}{\cosh \xi_2}+
\tanh \rho x_1\sinh \rho x_2 \, A \right)  }  , \\[10pt]
\displaystyle{ \{x_1,\xi_2\} = - z \cosh \xi_2 \,B  ,\qquad
\{x_2,\xi_2\} = z\left(
 \frac{\cosh \rho x_1}{\cosh \rho x_2}\,\cosh \xi_1 - \cosh \xi_2 
\right)    }  ,\\[12pt]
\displaystyle{\{x_2,\xi_1\} = -z A  ,\qquad    \{\xi_1,\xi_2\} = z\rho 
\sinh \rho x_1 \left(C-
\frac{\tanh \xi_2 }{\cosh^2\! \rho x_2}  \right) . }  
\end{array}
\label{gd}
\ee 

\noindent
$\bullet$ The remaining ones:
\be  
\begin{array}{l} 
\displaystyle{\{x_0,\theta\} = -  \frac{z}{\cosh \rho x_1 }\, B ,
 \qquad \{x_0,\xi_1\} = z\left(
\frac{\sinh \xi_2}{ \cosh \rho x_1}\,B-  \frac{\sinh \xi_1\cosh
\xi_2}{ \cosh \rho x_1\cosh \rho x_2}  \right) } , \\[12pt]
\displaystyle{  \{x_0,\xi_2\}
= - z C  , \qquad  \{\theta,x_1\} =   z \, \frac{\cosh \rho x_1}{\cosh
\xi_2}\, C  , \qquad
 \{\theta,x_2\} = - z \,\frac{\cosh \rho x_1\sinh
\xi_1}{\cosh \rho x_2\cosh
\xi_2}   } , \\[10pt]
\displaystyle{ \{\theta,\xi_1\}  = -z\rho \left(
\tanh \rho x_2+ \tanh \rho x_1  \,B  \right)  ,\qquad 
\{\theta,\xi_2\} =  \frac{z\rho \sinh \rho x_1 }{\cosh^2\! \rho x_2\cosh
\xi_2}   }  ,
\end{array}
\label{ge}
\ee 
where the functions $A$, $B$ and $C$ are  
\be  
\begin{array}{l} 
\displaystyle{A=\frac {\sinh \rho x_1\sinh \rho
x_2+ \cosh \rho x_1\sinh \xi_1\tanh \xi_2} {\cosh \rho x_2}  }  , \\[12pt]
\displaystyle{B=\frac {\sinh \rho x_1\tanh \rho x_2 \cosh \xi_1+  \sinh
\xi_1\sinh \xi_2} {\cosh \rho x_2\cosh
\xi_2}    } ,
\\[12pt]
\displaystyle{C=\frac {\sinh \rho
x_1\tanh \rho x_2 \sinh \xi_1+  \cosh \xi_1\sinh \xi_2}{\cosh \rho x_1\cosh \rho x_2} . }
 \end{array}
\label{gf}
\ee

Notice that the   commutators (\ref{cd}) are recovered
from (\ref{gc})--(\ref{gf}) by taking the  first-order in the group
coordinates.

We remark that any other choice for the matrix
element of $SO_\k(2,2)$ would lead to other set  of vector fields 
formally different to those written in table \ref{table2}  and, therefore, it would give
rise to    Poisson--Lie brackets     different from (\ref{gc})--(\ref{gf}). However,
by construction, all of these possible vector fields and Poisson--Lie structures are
equivalent by means of changes of basis involving the  
group coordinates.


\section{Non-commutative  (anti-)de Sitter spaces}

In general, the simplest way to quantize Poisson--Lie
structures~\cite{Tak,Drlb}  consists in the usual Weyl substitution of
the initial  Poisson brackets between commutative coordinates by
commutators between non-commutative coordinates.  Several
quantum deformations of  {\em non-semisimple} groups have been
constructed by applying this procedure; amongst them we underline   
the $\kappa$-Poincar\'e group~\cite{Maslanka,Majid:1994cy,Zak,LukR}, for
which the full set of   commutation rules are   linear in the deformation
parameter.
 Nevertheless, we stress that a quantum group does not always coincide
with the Weyl quantization of their underlying Poisson--Lie
brackets, specially when dealing with semisimple groups as the 
(A)dS ones, since ordering problems often appear during the quantization
procedure.

In our case, the Poisson--Lie brackets (\ref{gc})--(\ref{gf}) are the commutative
counterpart of the  full non-commutative quantum (A)dS groups ${\rm Fun}_z(SO_\k(2,2))$.
Thus if we write a generic Poisson--Lie bracket as  
$$
\{  y^i,  y^j\}= z f( y^1,\dots,y^6 ) ,
$$
 its non-commutative version  would read  
$$
[\hat
y^i,\hat y^j]=z f( \hat y^1,\dots,\hat y^6 ) +o(z^2) ,
$$
whose first-order in all $\hat y^k$ and $z$ 
is given by (\ref{cd}), while the $o(z^2)$ terms come from the reordering of the 
 quantum coordinates
$\hat y^k$. By following this point of view,   we will analyse the
generalization of the first-order non-commuta\-tive spaces presented in section 3
to   higher orders in the quantum coordinates and up to second-order in $z$.


\subsection{Non-commutative   spacetimes}

The first set of   Poisson--Lie brackets (\ref{gc}) allows us to introduce the
defining commutation relations of the (2+1)D non-commutative (A)dS
spacetimes, namely  
\be  
\begin{array}{l} 
{\displaystyle{ [\hat x_0, \hat x_1] =-z\,\frac{\tanh\rho \hat x_1}{\rho
\cosh^2\rho
\hat x_2}+o(z^2)= }} -z \hat x_1 + \frac 13 z\k \hat x_1^3+ z\k\hat x_1\hat
x_2^2 +o(z^2)  , \\[10pt]
{\displaystyle{[\hat x_0,\hat x_2] =-z\,\frac{\tanh\rho \hat x_2}{\rho
}+o(z^2)=}}   -z \hat x_2+\frac 13 z \k \hat x_2^3+o(z^2)  ,\\[10pt]
 [\hat x_1,
\hat x_2] =0+o(z^2) .  
\end{array} 
\label{ha}
\ee  
These expressions  demonstrate  how the  underlying first-order  ${\bf M}_z^{2+1}$ (\ref{ea})
for the  (A)dS  groups is now generalized with an explicit dependence on
the   curvature $\k$, in such a way that   three different cases appear.  Furthermore, space
isotropy is preserved in the quantum case since $\hat \theta$ is again absent (as well
as the quantum boost  coordinates $\hat{ \boldsymbol{
\xi}}$).

It is worth
mentioning that the asymmetric form of   (\ref{ha}) 
could be expected fom the beginning (see, e.g., the classical metric (\ref{dh})) as
we are dealing with   local quantum coordinates. However, if we consider  
non-commutative ambient (Weierstrass) coordinates $(\hat \s_3,\hat \s_\mu)$
defined in terms of the former ones $\hat x_\mu$ by  the same formal relations
(\ref{dg}) and subjected to the constraint (\ref{dd}), we obtain the
non-commutative   spacetimes written in a fully symmetric way  as a quadratic algebra:
\be  
\begin{array}{l} 
[\hat \s_0, \hat  \s_i]=-z   \,\hat  \s_3\hat  \s_i+o(z^2) , \qquad 
[\hat  \s_1,\hat  \s_2]=0+o(z^2) , \\[2pt]  
[\hat  \s_3,\hat \s_0]=z w \,\hat \>\s^2+o(z^2) ,  \qquad\ \   
[\hat \s_3,\hat \s_i]=z w \, \hat \s_0\hat \s_i +o(z^2)  .
\end{array} 
\label{hb}
\ee  
These expressions are clearly invariant under the permutation $\hat \s_1\leftrightarrow \hat \s_2$. 
 The two first relations in (\ref{hb}) are just  ${\bf M}_z^{2+1}$, since $\hat 
\s_3\to 1 $ when $\k\to 0$.


\subsection{Non-commutative  spaces of worldlines}

Likewise, a first insight into the non-commutative (A)dS and
Minkowskian spaces of time-like lines can be performed by starting with the  
Poisson--Lie brackets    (\ref{gd}) among space  $\>x$ and boost ${
\boldsymbol{ \xi}}$ group coordinates.

Let us consider firstly the quantum
Poincar\'e group with $\k=0$. In this case, the expressions (\ref{gd}) are rather
simplified. Their corresponding quantum deformation reads
$$ 
\begin{array}{l} 
\displaystyle{ [\hat x_1,\hat  \xi_1] = z\left(
1-\frac{\cosh\hat  \xi_1}{\cosh\hat  \xi_2} \right)   , \qquad    
 [\hat x_2,\hat  \xi_2] = z\left(
 \cosh \hat \xi_1 - \cosh\hat  \xi_2  \right)  , }
\\[12pt]
\displaystyle{   [\hat x_1, \hat  \xi_2 ] = - z \sinh
\hat \xi_1  \sinh\hat  \xi_2  ,\qquad   [\hat x_2, \hat  \xi_1 ] = -z \sinh \hat
\xi_1\tanh \hat \xi_2  ,  } \\[6pt]
\displaystyle{     
[\hat x_1, \hat  x_2 ] =0,  \qquad  [\hat \xi_1, \hat
 \xi_2 ] =0  , }  
\end{array}
$$
which determine the   complete (in all orders in 
$z$,   $\hat
\>x$ and 
$\hat{ \boldsymbol{ \xi}}$) non-commutative Minkowskian space  of worldlines ${\bf
LM}_z^{2\times 2}$.  Hence the commutativity of    the first-order
  relations $[\hat x_i,\hat  \xi_j]$  is lost (see (\ref{ed})). It is worth mentioning that  
$[\hat \xi_1, \hat
\xi_2 ]=0$ ensures the self-consistency of the 
  remaining commutators as   no ordering problems appear.   Moreover, this condition
means that the 2D velocity space remains non-deformed (commutative).

On the contrary, in the (A)dS cases with $\k\ne 0$  the brackets  
(\ref{gd}) are rather involved,  again asymmetric, and  $\hat \xi_1, \hat \xi_2 $
do not  longer commute. As in the classical case, we can consider the 
non-commutative ambient (Pl\"ucker) coordinates
$(\hat \et_3, \hat {\boldsymbol{\et}}, \hat {\boldsymbol{\yy}},
\hat \yy_3)$   formally defined by (\ref{dn}), but now with
non-commutative entries  $\hat
\>x$ and   $\hat{ \boldsymbol{ \xi}}$. Then the non-commutative  (A)dS
  spaces of worldlines are given  by the following quadratic relations
\be  
\begin{array}{l} 
\displaystyle{ [\hat  \yy_1,\hat   \yy_2]=-z \left(\hat 
 \et_3-(\hat   \et_3^2-\k\hat  \>\yy^2)
\right)\hat   \yy_3 +o(z^2) ,  \qquad [\hat   \yy_3,\hat  \et_3]=0+o(z^2) , } \\[8pt]
\displaystyle{ [\hat  \et_1,\hat  \et_2]=-z\k\left(\hat 
\et_3-(\hat  \et_3^2-\k\hat  \>\yy^2)
\right)\left(\frac{\hat  \et_3-1}{\hat  \et_3}\right)\hat  \yy_3  +o(z^2)  , }
\\[12pt]
 \displaystyle{
[\hat  \et_3,\hat  \et_i]= z\k  \left(  \epsilon_{ij}\hat  \et_j
\hat  \yy_3 + \hat  \yy_i(\hat  \et_3^2-\k\hat  \>\yy^2)\left( 
\frac{\hat  \et_3-1}{\hat  \et_3}\right)
 \right)   +o(z^2)  , }\\[12pt]
\displaystyle{[\hat  \et_3,\hat  \yy_i]= z  \left(  \k\epsilon_{ij}\hat  \yy_j
\hat  \yy_3 - \hat  \et_i(\hat  \et_3^2-\k\hat  \>\yy^2)\left( 
\frac{\hat  \et_3-1}{\hat  \et_3}\right)
 \right)  +o(z^2) ,  }\\[12pt]
\displaystyle{
[\hat  \yy_3,\hat  \yy_i]= z  \left(  \epsilon_{ij}\hat  \yy_j
(\hat  \et_3-1) -\hat   \et_i(\hat  \et_3^2-\k\hat  \>\yy^2) 
\frac{\hat  \yy_3}{\hat  \et_3}  \right)   +o(z^2) ,  }\\[12pt]
\displaystyle{
[\hat  \yy_3,\hat  \et_i]= z  \left(  \epsilon_{ij}\hat  \et_j
(\hat  \et_3-1) +\k \hat  \yy_i(\hat  \et_3^2-\k\hat  \>\yy^2) 
\frac{\hat  \yy_3}{\hat  \et_3}  \right)    +o(z^2),  } \\[12pt]
\displaystyle{
[\hat  \et_i,\hat  \yy_j]= z \delta_{ij} \hat  \et_3 \left(\hat 
\et_3-1+\k\hat  \yy_3^2
\right) -z\hat  \et_i\hat  \et_j\hat \et_3+z\k 
\hat  \yy_i\hat  \yy_j\left(\k\frac{\hat  \yy_3^2}{\hat  \et_3}-1
\right)    }\\[12pt]
\displaystyle{\qquad\qquad\qquad
+z\k \left( \epsilon_{ik}\hat  \et_k\hat   \yy_j\left(\frac{\hat  \et_3-1}{\hat  \et_3}\
\right)  +\epsilon_{jk}\hat  \et_k\hat   \yy_i \right) \hat  \yy_3  +o(z^2) ,}
\end{array}
\label{hd}
\ee 
which are invariant under the map defined by
 \be
(\hat \et_3,\hat \et_1,\hat \et_2,\hat \yy_1, \hat \yy_2,  \hat \yy_3)\to (\hat \et_3 ,\hat
\et_2 ,\hat
\et_1 ,\hat
\yy_2,\hat \yy_1, -\hat \yy_3 )  .
\label{he}
\ee

The flat contraction $\k=0$ of (\ref{hd}) to ${\bf
LM}_z^{2\times 2}$   yields
$$ 
\begin{array}{l} 
\displaystyle{ [\hat  \yy_1,\hat   \yy_2]= z \hat   \yy_3\hat 
 \et_3 \left( \hat   \et_3  -1
\right)  ,  \qquad   [\hat   \yy_3,\hat  \et_3]=0  ,\qquad  
 [\hat  \et_1,\hat  \et_2]=0    ,\qquad  
[\hat  \et_3,\hat  \et_i]= 0 ,  }\\[12pt]
\displaystyle{[\hat  \et_3,\hat  \yy_i]= -z   \hat  \et_i \hat 
\et_3   \left( 
 {\hat  \et_3-1} 
 \right)  ,   \qquad 
[\hat  \yy_3,\hat  \yy_i]= z     \epsilon_{ij}\hat  \yy_j
(\hat  \et_3-1) -z \hat   \et_i \hat  \et_3  
 \hat  \yy_3    +o(z^2)   , }\\[12pt]
\displaystyle{
[\hat  \yy_3,\hat  \et_i]= z    \epsilon_{ij}\hat  \et_j
(\hat  \et_3-1)      ,  \qquad\ \ \,
[\hat  \et_i,\hat  \yy_j]= z \delta_{ij} \hat  \et_3 \left(\hat 
\et_3-1 
\right) -z\hat  \et_i\hat  \et_j\hat \et_3   .}
\end{array}
$$
The commutativity of the 2D velocity space in ${\bf
LM}_z^{2\times 2}$ now comes
from the fact that the commutators involving the ``momentum-like"
coordinates $(\hat
\et_3, \hat {\boldsymbol{\et}})$ vanish; note that they only depend on  $\hat{ \boldsymbol{ \xi}}$, which commute in this case. On the contrary, the ``position-like" ones  $(\hat
\yy_3, \hat {\boldsymbol{\yy}})$ do not commute; they   depend on both  $\hat{ \boldsymbol{ \xi}}$  and $\hat{ \bf{ x}}$ (see  (\ref{do})). We     stress that
this quantum Poincar\'e group property is in full agreement with the study
developed in
\cite{Bruno:primo,Bruno:2002wc} by working with the (dual) quantum algebra. In these
works it is shown how the $\kappa$-deformed Poincar\'e boost  transformations  close a 
group as in the non-deformed case, and the additivity of the boost parameter for
transformations  along a same direction is also preserved. Therefore, the   relations
(\ref{hd})  
 indicate that such properties may be either lost or somewhat modified in the quantum
(A)dS groups.


\section{Quantum (anti-)de Sitter algebras}

In this section we firstly review the Hopf algebra structure  and the associated
invariants of the quantum (A)dS algebras that quantize  the  
cocommutators (\ref{cb}), 
  commutators (\ref{ba}) and Casimirs (\ref{bc}). These results are presented in the
kinematical basis in which they were formerly obtained~\cite{CK3}, and   the contraction
$\k=0$ gives the $\kappa$-Poincar\'e written in the form deduced in~\cite{Giller}.  We
also point out some remarks concerning the connection between these structures  and
quantum gravity that has been   introduced in~\cite{amel}.
Secondly, we obtain a new basis
through a non-linear map for these quantum algebras in such a manner that    the
$\kappa$-Poincar\'e algebra is recovered in the so called bicrossproduct
basis~\cite{Majid,Majid:1994cy}. This change of basis, at the level of the quantum
algebra, is the dual counterpart of a change of non-commutative coordinates on the
quantum group.


\subsection{``Symmetrical" basis}

The Drinfel'd--Jimbo quantum deformation of $so(4,\mathbb{C})$ was obtained
in~\cite{Ita}     by considering   two copies of the quantum
$sl(2)$ algebra~\cite{ drinfeld87}  and applying the prescription 
$U_z(so(4,\mathbb{C}))=U_z(sl(2))\oplus U_{-z}(sl(2))$. From this result, a
further analysis of   real forms together with a contraction scheme led to 
quantum deformations of the family of (2+1)D   kinematical
algebras~\cite{CK3,LBC}  which included, amongst others, the  $\kappa$-deformation 
of the three relativistic algebras (\ref{ba}) with underlying Lie bialgebra  
(\ref{cb}), that we denote here by $U_z(so_\k(2,2))$.   The Hopf structure of
$U_z(so_\k(2,2))$ is characterized by the following coproduct  
 and commutation relations~\cite{CK3}:
\be 
\begin{array}{l} 
 \Delta(P_0)=1\otimes P_0+P_0\otimes 1, \qquad
\Delta(J)=1\otimes J+J\otimes 1 ,\\[6pt]
 \Delta(P_i)={\rm e}^{-\frac z2 P_0 }\cosh(\frac z2 \rho J)\otimes P_i+
P_i \otimes {\rm e}^{\frac z2  P_0 }\cosh(\frac z2 \rho J)\\[4pt]
\qquad\qquad +  \rho\, {\rm e}^{-\frac z2 P_0 }\sinh(\frac z2 \rho
J)\otimes \epsilon_{ij} K_j-  \rho\,  \epsilon_{ij} K_j \otimes {\rm e}^{\frac z2 
P_0 }\sinh(\frac z2 \rho J) ,\\[6pt]
\Delta(K_i)={\rm e}^{-\frac z2 P_0 }\cosh(\frac z2 \rho J)\otimes K_i+
K_i \otimes {\rm e}^{\frac z2  P_0 }\cosh(\frac z2 \rho J)\\[4pt]
\qquad\qquad 
\displaystyle{ -   {\rm e}^{-\frac z2 P_0 }\, \frac{ \sinh(\frac z2 \rho
J) }{\rho}\otimes \epsilon_{ij} P_j+  \epsilon_{ij} P_j \otimes {\rm e}^{\frac z2 
P_0 }  \, \frac{ \sinh(\frac z2 \rho
J) }{\rho} } ,
\end{array}
\label{fa}
\ee 
\be  
\begin{array}{l} 
[J,P_i]=   \epsilon_{ij}P_j , \qquad
[J,K_i]=   \epsilon_{ij}K_j  , \qquad\   [J,P_0]= 0 ,  \\[2pt]
\displaystyle {[P_i,K_j]=-\delta_{ij}\frac{\sinh (zP_0)}{z}\,\cosh(z\rho J)}
, \qquad [P_0,K_i]=-P_i  ,\qquad\ 
[P_0,P_i]=\k K_i   , \\[2pt]
\displaystyle {  [P_1,P_2]= -\k \cosh (zP_0)  \, \frac{
\sinh(z\rho J) }{z \rho} },    \qquad  \displaystyle { [K_1,K_2]= -\cosh (zP_0)\, 
\frac{\sinh(z\rho J) }{z \rho} }  .  
\end{array}
\label{fb}
\ee 
 Counit and antipode maps   can directly be derived
from the Hopf algebra axioms.  The deformation of the two Casimir invariants  
(\ref{bc}) turns out to be
\be  
\begin{array}{l} 
{\cal C}=4 \cos (z\rho) \left\{   {\displaystyle { \frac{\sinh^2(\frac z 2
P_0)}{z^2} }} \, \cosh^2\left(\frac z 2\rho J\right)  +
{\displaystyle {  \frac{\sinh^2(\frac z 2
\rho J)}{z^2}}}  \, \cosh^2\left(\frac z 2 P_0\right)
\right\}   \\[12pt] 
\displaystyle {\quad\qquad\  
- \frac{\sin (z\rho)}{z\rho} \left(\>P^2+\k \>K^2 \right)  }, \\[6pt]
\displaystyle {{\cal W}= -\cos (z\rho)\, \frac{\sinh(z\rho J)}{z\rho}\, 
\frac{\sinh(z P_0)}{z}+\frac{\sin (z\rho)}{z\rho} (K_1P_2-K_2P_1 )} .
\end{array}
\label{fc}
\ee 

This deformation  is governed by the generators spanning the  
isotropy subgroup of a worldline, $P_0$ and $J$, which remain   
 non-deformed at the level of the coproduct (\ref{fa}). Thus  deformed functions of
$P_0$ and $J$ arise  for   the coproduct of  $\>P$ and  $\>K$ in both spaces in   the
tensor product, in such a manner that their  coproduct is invariant under the
composition
$\sigma\circ{\cal T}$ of the flip operator, $\sigma(x\otimes y)=y\otimes x$, and a
``parity" operator ${\cal T}$ acting on the deformation parameter as ${\cal T}(z)=-z$.
Hence we    say that $U_z(so_\k(2,2))$ is written in a ``symmetrical" basis.

 The physical dimension  of    $z$ is inherited from $P_0$, $[z]=[P_0]^{-1}$, so that 
this can be interpreted as a fundamental length (provided that $c=1$), which in the
usual DSR theories is considered to be of the order of  the Planck length $l_p$.

Expressions  (\ref{fa})--(\ref{fc})  show the commutativity character of the
l.h.s.\ of the diagram (\ref{aa}); the  limit $z\to 0$ in each of the three
particular quantum algebras contained in the family $U_z(so_\k(2,2))$ leads to
the corresponding Lie algebra $so_\k(2,2)$ and its invariants, while the 
contraction  
$\k= 0$ ($\rho\to 0$) in $U_z(so_\k(2,2))$ gives rise to the
$\kappa$-Poincar\'e algebra and its deformed invariants in the basis  formerly
worked out in~\cite{Giller}. Both types of limits can be applied {\em separately}.
Such a viewpoint suggests some kind of ``duality" between quantum deformations
induced by $z$ and  ``curvature deformations" parametrized by $\k$; recall that (A)dS symmetries can be considered as a ``classical" deformation of  Poincar\'e invariance~\cite{Ober,Ober2,Gilmore}. 

The dual relationship  between quantum deformation
parameters and classical deformation ones  ($z\leftrightarrow \k$)  was already
analysed for the $(1+1)$D case in~\cite{Poisson} and, in fact,   the role of
$z$ as a curvature also arises within the Drinfel'd-double approach, as commented in
section 3.1 (see~\cite{Kowalskia,Kowalskib}).
Moreover,  the ``semidualization" approach in 2+1 quantum gravity introduced in~\cite{MSsemid} (see also~\cite{OSsemid, Oseir}) provides a more complete Hopf algebraic understanding of this duality between the Planck scale $l_p$ and the cosmological constant $\k=\ro^2=-\Lambda$, and shows its direct connection with the quantum version of the so-called Born reciprocity principle~\cite{Born} between (now non-commutative) coordinates and (curved) momenta. This framework is also helpful in order to understand the existing constraints on contraction limits involving these two parameters, that we will discuss in the sequel.


\subsection{Contractions and the Planck length}

It has been shown   (see~\cite{amel} and
references therein) that there exists a natural link bewteen the  deformed commutation
relations (\ref{fb}) for the (A)dS algebras   with $\k\ne 0$  and
2+1 quantum gravity. In the latter framework, the curvature   and
deformation parameters  can be identified with the cosmological constant
$\Lambda\equiv - \k$ and the fundamental Planck length $l_p\equiv z$.
Hence, by considering these identifications, the above results show that both limits
$\Lambda\to 0$ and 
$l_p\to 0$ can be taken independently (if $\Lambda\to 0$ we
obtain $\kappa$-Poincar\'e with $l_p\equiv z$), and that a  ``duality"
$l_p\leftrightarrow
\Lambda$ might exist.  
Nevertheless, in~\cite{amel}  it is emphasized that the contraction to
$\kappa$-Poincar\'e should be taken as the {\em simultaneous} limits $z\to 0$ and
$\Lambda l_p^2\to 0$, due to the  coupling   $z=
\sqrt{\Lambda}l_p$ existing in 2+1 quantum gravity. This fact can be
explained from a Lie bialgebra contraction   approach~\cite{LBC} as follows.

Alternatively to the expressions (\ref{fa}) and (\ref{fb}) defining
$U_z(so_\k(2,2))$, that unify the Hopf structure for the quantum (A)dS
and Poincar\'e  algebras parametrized by $\k$, one could
have started from the quantum AdS algebra
$U_z(so(2,2))$ with $\k=1$ (or from $U_z(so(3,1))$ with $\k=-1$). Next the Lie bialgebra contraction
analysis of the $r$-matrix (\ref{ca}) and Lie bialgebra (\ref{cb}) shows that
there exists a unique quantum In\"on\"u--Wigner contraction (a coboundary Lie bialgebra contraction) 
that ensures the convergency of both (\ref{ca}) and  (\ref{cb}). This is
defined through the   new generators $Y'$ and deformation parameter $z'$ given 
by~\cite{CK3, LBC}:
\be
P'_0=\sqrt{\k} P_0  ,\qquad P'_i=\sqrt{\k} P_i ,\qquad  K'_i= K_i ,\qquad J'=J  ,
\qquad z'=z/\sqrt{\k} ,
\label{ffd}
\ee
($\sqrt{\k}=\rho$ plays the role of the usual In\"on\"u--Wigner
contraction parameter). This map is associated to     the
$\mathbb{Z}_2$-grading of
$U_z(so(2,2))$ given by the composition of the deformed parity and time-reversal
(\ref{cf}): 
$$
\Pi_z\Theta_z  :  (P_0, \>P , \>K,J; \ z)\to  (-P_0,- \>P, 
\>K,J; \ -z) .
$$
By computing   the coproduct and commutation
relations for the new generators $Y'$ from $U_z(so(2,2))$ and then by
applying the limit 
$\sqrt{\k}\to 0$, one finds the $\kappa$-Poincar\'e algebra $U_{z'}(iso(2,1))$ in
terms of the transformed deformation parameter $z'$. Therefore   if $z'$ is
{now} taken as the Planck length $l_p$ and $\k\equiv - \Lambda>0$,    the map
(\ref{ffd}) means that
$z=\sqrt{-\Lambda}l_p$, so that the contraction $\sqrt{-\Lambda}\to 0$  conveys the
limit $z\to 0$ as well, leaving $z'\equiv l_p$ as the only fundamental scale in
$\kappa$-Poincar\'e, in agreement with~\cite{amel}.


\subsection{``Bicrossproduct-type" basis}

So far, most of the applications of the $\kappa$-Poincar\'e algebra have been developed
by working in the so called bicrossproduct basis~\cite{Majid:1994cy,LukNR}, in
which the Lorentz sector has non-deformed commutation rules. Thus it is natural to
wonder about the existence of a non-zero curvature counterpart of such a basis. This can
be achieved by means of the $\k$-generalization of the invertible non-linear map
introduced for $\kappa$-Poincar\'e in~\cite{Majid:1994cy}:
\be
\begin{array}{l} 
   \tilde P_0=P_0 , \qquad \tilde J= J , \\[4pt] 
 \tilde P_i={\rm e}^{-\frac z2 P_0 }\left\{ \cosh(\frac z2 \rho
J) P_i-\rho    \sinh(\frac z2 \rho
J)    \,  \epsilon_{ij} K_j  \right\}  , \\[6pt] 
 \tilde K_i={\rm e}^{-\frac z2 P_0 }\left\{ \cosh(\frac z2 \rho
J) K_i+ \displaystyle { \frac{ \sinh(\frac z2 \rho
J) }{\rho}  \,  \epsilon_{ij} P_j }\right\}  .
\end{array}
\label{xa}
\ee
This transformation can be written as the following matrix transformation of $\>P$ and
$\>K$ depending on functions of the  generators of the isotropy subgroup of a worldline:
$$
\left(\begin{array}{c}
\tilde P_1\cr 
\tilde P_2\cr 
\tilde K_1\cr 
\tilde K_2
\end{array}\right) 
=\exp\left\{ -\frac{z}{2} P_0 \left(\begin{array}{cccc}
1&0&0&0\cr 
0&1&0&0\cr 
0&0&1&0\cr 
0&0&0&1
\end{array}\right) 
 \right\} 
\exp\left\{ -\frac{z}{2}\, J \left(\begin{array}{cccc}
0&0&0&w\cr 
0&0&-w&0\cr 
0&-1&0&0\cr 
1&0&0&0
\end{array}\right) 
 \right\}
\left(\begin{array}{c}
 P_1\cr 
  P_2\cr 
  K_1\cr 
  K_2
\end{array}\right) .
$$

The transformed coproduct and commutation rules of  $U_z(so_\k(2,2))$ in this
new basis turn out to be
 \be 
\begin{array}{l} 
 \Delta(\tilde P_0)=1\otimes \tilde P_0+\tilde P_0\otimes 1 , \qquad
\Delta(\tilde J)=1\otimes \tilde J+\tilde J\otimes 1 , \\[6pt]
 \Delta(\tilde P_i)={\rm e}^{-z \tilde P_0 } \otimes \tilde P_i+
\tilde P_i \otimes  \cosh(z \rho \tilde J) -  \rho  \epsilon_{ij} \tilde K_j
\otimes  \sinh(z \rho \tilde J)  , \\[4pt]
 \Delta(\tilde K_i)={\rm e}^{-z \tilde P_0 } \otimes \tilde K_i+
\tilde K_i \otimes  \cosh(z \rho \tilde J) +  \epsilon_{ij}\tilde  P_j \otimes
\displaystyle{\frac{ \sinh(z \rho \tilde J) }{\rho} } ,
\end{array}
\label{fe}
\ee 
\be
\begin{array}{ll} 
[\tilde J,\tilde P_i]=   \epsilon_{ij}\tilde P_j  ,&\qquad
[\tilde J,\tilde K_i]=   \epsilon_{ij}\tilde K_j  , \qquad\quad   [\tilde J,\tilde
P_0]= 0    , \qquad\quad [\tilde P_0,\tilde K_i]=-\tilde P_i , \\[4pt]
[\tilde P_0,\tilde P_i]=\k \tilde K_i  ,  &\qquad \displaystyle {  [\tilde
P_1,\tilde P_2]= -\k  \, \frac{
\sinh(2z\rho \tilde J) }{2 z  \rho} }  ,  \qquad  \displaystyle { [\tilde
K_1,\tilde K_2]= - \frac{\sinh(2 z\rho \tilde J) }{2 z \rho} }  ,  \\[4pt]
\multicolumn2{l}{ \displaystyle {[\tilde P_i,\tilde K_j]= \delta_{ij}\left\{
 \frac{{\rm e}^{-2 z\tilde P_0} - \cosh(2 z\rho\tilde  J)}{2z}- 
\frac{\tan (z\rho)}{ 2\rho} \left( \>{\tilde P}^2+\k \>{\tilde K}^2\right)
\right\}   }   }\\[8pt]
\multicolumn2{l}{ \displaystyle {\qquad\qquad\qquad   +\frac{\tan (z\rho)}{ 
\rho}   (\tilde P_j\tilde P_i+\k\tilde  K_i\tilde  K_j)  }.  } 
\end{array}
\label{fe2}
\ee
 Therefore   the Lorentz sector remains deformed in the quantum (A)dS
algebras with $\k\ne 0$, while the contraction $\k=0$ ($\rho\to 0$) produces the  
$\kappa$-Poincar\'e algebra in the bicrossproduct basis, in which the only deformed
commutation rules are   $[\tilde P_i,\tilde K_j]$.   However,   although the Lorentz
sector in  $\kappa$-Poincar\'e is non-deformed, the quantum  deformation is still  kept
in  the coproduct for the boost generators (\ref{fe}).  Note also that this new
coproduct is not invariant under the   map
$\sigma\circ{\cal T}$.

The corresponding deformed Casimirs are obtained from (\ref{fc}) and read
\be  
\begin{array}{l} 
{\cal C}=4 \cos (z\rho) \left\{   {\displaystyle { \frac{\sinh^2(\frac z 2
\tilde P_0)}{z^2} }} \, \cosh^2\left(\frac z 2\rho \tilde J\right)  +
{\displaystyle {  \frac{\sinh^2(\frac z 2
\rho \tilde J)}{z^2}}}  \, \cosh^2\left(\frac z 2 \tilde P_0\right)
\right\}   \\[12pt] 
\displaystyle { \qquad\  
- \frac{\sin (z\rho)}{z\rho} \,{\rm e}^{z\tilde P_0}\left\{ \cosh (  z
 \rho\tilde  J )  \left(\>{\tilde P}^2+\k \>{\tilde K}^2
\right)-2\rho \sinh (z \rho\tilde  J ) (\tilde K_1\tilde P_2-\tilde K_2\tilde P_1
)  \right\} } , \\[10pt]
\displaystyle {  {\cal W}= -\cos (z\rho)\, \frac{\sinh(z\rho\tilde  J)}{z\rho}\, 
\frac{\sinh(z\tilde  P_0)}{z} }\\[10pt]
\displaystyle {\qquad\   +\frac{\sin (z\rho)}{z\rho} \,{\rm e}^{z
\tilde P_0}\left\{ 
\cosh (z \rho  \tilde  J ) (\tilde K_1\tilde P_2-\tilde K_2\tilde P_1 )
-\frac{\sinh (z
\rho\tilde  J )} {2 \rho}
\left(  \>{\tilde P}^2+\k \>{\tilde K}^2 \right) \right\} }.
\end{array}
\label{fc2}
\ee

We remark that for both bases the full Hopf structure  $U_z(so_\k(2,2))$  is
invariant under the quantum involutions (\ref{cf}) (as it should be) and also  under the
``classical" symmetry
\be
(P_0,P_1,P_2,K_1,K_2, J)\to (P_0,P_2,P_1,K_2,K_1, -J)  ,
\label{sym}
\ee
which shows that the equivalences $P_1\leftrightarrow P_2$ and 
 $K_1\leftrightarrow K_2$ are preserved by this quantum deformation. This, in turn,
means that there is no privileged space/boost direction. 
The   analogous (dual) maps  to (\ref{sym}) in the non-commutative spacetimes and spaces
of worldlines are given by  the interchange $\hat \s_1\leftrightarrow \hat \s_2$    and 
(\ref{he}), respectively.


\section{Concluding remarks}

We have presented a  unified and
global study of the $\kappa$-deforma\-tion of the (A)dS
and Poincar\'e algebras and groups in 2+1 dimensions by making use of an
explicit contraction parameter
$\k$ that corresponds to the curvature/cosmological constant of the underlying
classical spacetimes.    We remark that the limit
$\k\to 0$ is always well defined in all the
expressions, thus providing a straightforward Poincar\'e/Minkowskian counterpart of all the results here presented.

 At the quantum algebra level,  we
have    introduced a new basis (\ref{xa}) for  the quantum (A)dS algebras,
which is the analogous of the $\kappa$-Poincar\'e bicrossproduct basis.
As far as the (dual) quantum groups are concerned, the   results    cover the
non-commutative (A)dS spacetimes (\ref{ha}) up to second-order in the deformation
parameter, thus   generalizing the  $\kappa$-Minkowskian spacetime  (\ref{ea}) which turns out to
be   their common first-order seed.  Furthermore, we have presented
  the first   approach,  to the best of our knowledge,    to non-commutative spaces
of worldlines. For the three relativistic cases they  are
related to  phase spaces   $(\hat \>x, \hat{ \boldsymbol{ \xi}}) \leftrightarrow  
(\>P,\>K)$, which  are different from the  $\kappa$-Poincar\'e phase
spaces proposed in~\cite{Kowalskia,Kowalskib} (see also \cite{Kowalskic}).  
 We also remark that an
appropriate treatment of the classical (non-deformed) structures have allowed us to write
both non-commutative spacetimes and spaces of worldlines in terms of ambient (``global")
coordinates, by starting with the particular expressions   (\ref{dg}) and (\ref{dn})  for the (``local") parametrizations of 
such spaces. In such ambient coordinates, the non-commutative spaces turn out to be determined through quadratic commutation relations (\ref{hb}) and  (\ref{hd}). Morevover, they are shown to be
symmetric with respect to some maps that generalize the $P_1\leftrightarrow P_2$ and $K_1\leftrightarrow K_2$ invariance 
 of the dual   quantum
algebras.

It is worth stressing that the connection between the Poisson--Lie group approach here presented and the role that classical $r$-matrices and Drinfel'd-doubles play  in the 
context of 2+1  quantum  gravity~\cite{AMII, FR, FRb, AT, Witten1, cm1, cm2, bernd1,bernd1b} has been studied in detail in the works~\cite{BHMplb, BHMCQG, BHMNsigma, BHMNplb, BMNphs,BGSHpoinc}. 
Also, the deformed Casimir operators (\ref{fc}) (or (\ref{fc2}))  can be used to provide modified dispersion relations, which should be related to those appearing in several phenomenological approaches to quantum gravity (see~\cite{AEN, Amelinodispersion1, Amelinodispersion1b, Mattingly}).
 
Some comments on other possible quantum deformations of the (A)dS algebras   are in order.
Since any possible quantum deformation of the (A)dS
algebras has to come from a classical $r$-matrix,   a
complete classification of the latter would be
certainly useful in order  to obtain and analyse  other physically interesting quantum
(A)dS groups  (for each of them, the vector fields displayed in table \ref{table2} would
give rise  to the associated Poisson non-commutative spaces). In this respect, the full classification of classical $r$-matrices for the (3+1)D Poincar\'e Lie algebra can be found in~\cite{Zakr}. For the (2+1)D (A)dS algebras a similar result has been recently given in~\cite{LukiBorowiec} (see also~\cite{tallin}).

On the other hand, the generalization of the (2+1)D $\kappa$-(A)dS algebras and groups to the (3+1)D case can be obtained by generalizing the  classical $r$-matrix (\ref{ca}) to (see~\cite{LBC})
\be
r=z(K_1\wedge P_1+K_2\wedge P_2+K_3\wedge P_3) +z\sqrt{\k}\, J_1\wedge J_2  ,
\label{za}
\ee
which include a term $J_1\wedge J_2 $ coming from the rotation sector. This term does not
appear
  either  in the
(3+1)D $\kappa$-Poincar\'e algebra (with $\k=0)$  or in  the (2+1)D $\kappa$-(A)dS algebras. 
A twisted version of (\ref{za}) with a second deformation parameter $\vartheta$ was considered in~\cite{praga} by imposing some physical requirements, and the very same classical $r$-matrix has been derived in~\cite{BHNplb} from a Drinfel'd-double approach, namely
$$
r =    z(K_{1}\wedge P_{1}+K_{2}\wedge P_{2}+K_{3}\wedge P_{3}) +z\sqrt{\k}\, J_1\wedge J_2+\vartheta  J_3 \wedge P_0 .
$$
The corresponding Poisson--Hopf algebra    has been recently constructed in~\cite{BHMNkappa} so obtaining the (3+1)D Poisson version of the coproduct   (\ref{fe}), commutation relations (\ref{fe2}) and Casimirs (\ref{fc2}). The associated  (3+1)D non-commutative spacetime, generalizing (\ref{ha}), is currently under investigation.

Finally,  we recall that   other different quantum (A)dS deformations    have been obtained by
  working in a {conformal basis} instead of a pure {kinematical} one, for instance, 
$so(3,2)$~\cite{herranz97,Mozb} and $so(4,2)$~\cite{Brunoc,Moza, Aizawa, vulpi},  and the Drinfel'd--Jimbo quantum AdS space at roots of unity has been worked out in~\cite{Stein}. Also, other fuzzy~\cite{JS,BM} and covariant~\cite{VH} non-commutative (A)dS spacetimes have also been  constructed.


\section*{Competing interests}

The authors declare that they have no competing interests.


\section*{Acknowledgments}

 This work has been partially supported by Ministerio de Econom\'{i}a y Competitividad (MINECO, Spain) under grants MTM2013-43820-P and   MTM2016-79639-P (AEI/FEDER, UE), by Junta de Castilla y Le\'on (Spain) under grants BU278U14 and VA057U16 and by the Action MP1405 QSPACE from the European Cooperation in Science and Technology (COST).


\end{document}